\newcommand{\nc}{\newcommand}
\nc{\rnc}{\renewcommand}
\nc{\CY}{Calabi-Yau}
\nc{\CYM}{Calabi-Yau manifold}
\nc{\CYMs}{Calabi-Yau manifolds}
\nc{\DB}{D-Brane}
\nc{\DBs}{D-Branes}
\nc{\SUSY}{supersymmetry}
\nc{\Kah}{K\"ahler}
\nc{\cs}{complex structure}
\nc{\beq}{\begin{equation}}
\nc{\eeq}{\end{equation}}
\nc{\ntwo}{${\cal N}=2$}
\nc{\nOne}{${\cal N}=1$}
\nc{\hs}{\hspace{0.2in}}
\nc{\Z}{{\mathbb Z}}
\rnc{\P}{{\mathbb P}}
\nc{\R}{{\mathbb R}}
\nc{\C}{{\mathbb C}}
\nc{\WP}{\mathbb{WP}}
\nc{\slag}{special Lagrangian}
\nc{\cn}{\C^n}
\nc{\rn}{\R^n}
\nc{\M}{{\cal M}}
\nc{\W}{{\cal W}}
\nc{\linefour}{{\cal O}_{\P^4}(-5)}
\nc{\linen}{{\cal O}_{\P^{n-1}}(-n)}

\documentclass[12pt]{article}
\usepackage{epsfig,amsmath,amsfonts,amssymb}
\begin{document}
\rightline{\vbox{\baselineskip12pt\hbox{CITUSC/02-008,
USC-02/02}\hbox{hep-th/0203266}\hbox{28 March 2002}}}
\vskip 1cm
\begin{center}
{\bf \Large \bf A Geometrical Construction of Rational Boundary States in Linear Sigma Models}
\vskip 1cm
{\large Kristian D. Kennaway} \\
\vskip .4cm
{\it Department of Physics and Astronomy}\\
{\it and}\\
{\it CIT-USC Center for Theoretical Physics}\\
{\it University of Southern California}\\
{\it Los Angeles, CA 90089-0484, USA}
\end{center}

\begin{abstract}
  Starting from the geometrical construction of special Lagrangian
  submanifolds of a toric variety, we identify a certain subclass of
  A-type D-branes in the linear sigma model for a \CYM\ and its mirror
  with the A- and B-type Recknagel-Schomerus boundary states of the
  Gepner model, by reproducing topological properties such as their
  labeling, intersection, and the relationships that exist in the
  homology lattice of the D-branes.  In the non-linear sigma model
  phase these special Lagrangians reproduce an old construction of
  3-cycles relevant for computing periods of the \CY, and provide
  insight into other results in the literature on special Lagrangian
  submanifolds on compact \CYMs.  The geometrical construction of
  rational boundary states suggests several ways in which new Gepner
  model boundary states may be constructed.
\end{abstract}

\section{Introduction}

The study of string theory on Calabi-Yau manifolds provides a
potential point of contact with low energy (``real-world'') physics.
One can obtain four-dimensional gauge theories with \nOne\ 
supersymmetry (which is usually desirable to maintain a degree of
computational control over the theory, as well as for phenomenological
reasons) by compactifying heterotic string theory on $\R^{3,1} \times
CY_3$ where $CY_3$ is a Calabi-Yau 3-fold: this point of view was
studied extensively in the 1980s and 1990s.  Since the discovery of
D-branes as extended objects carrying gauge fields on their
world-volume, one may also obtain four-dimensional gauge theories by
considering Type II string theory on $\R^{3,1} \times CY_3$ in the
presence of D-branes.  In order to obtain a four-dimensional gauge
theory with \nOne\ supersymmetry, these D-branes must be BPS.  BPS
D-branes come in two types, labeled A- and B-type according to which
type of supersymmetry is preserved.

The linear sigma model (LSM) is a useful tool for studying the phase
structure that exists in the \Kah\ moduli space of string theory
compactified on \CYMs .  Although the LSM is not itself a
conformally-invariant theory, it flows to the desired conformal field
theories in the infrared via renormalization group flow.  Typically
the \Kah\ moduli space phase structure contains a geometrical phase
where the infrared CFT is described by string theory on a \CYM, as
well as one or more ``non-geometrical'' phases where the IR CFT does
not have an obvious geometrical description, but is instead described
by an abstract CFT such as the IR limit of a Landau-Ginzburg (LG)
theory (in a certain limit of \Kah\ moduli space this is a Gepner
model \cite{gepner, gepner2}, an exactly solvable CFT), or hybrid
phases such as a LG theory fibred over a geometrical base space.

In a geometrical phase, A-type D-branes correspond to flat vector
bundles over \slag\ 3-cycles, while B-type D-branes correspond to
stable holomorphic vector bundles over holomorphic (even-dimensional)
cycles \cite{chernsimons, bbs, ooy}.  A similar A-/B-type
classification of D-branes exists in Gepner models, as studied in
\cite{rs, bstates}.  D-branes were studied in the linear sigma model in
\cite{dmirror, lgcft, dglsm}.  The spectrum of BPS D-branes for a
given Calabi-Yau has been studied in \cite{quintic, delliptic, k3fib,
  bpsnoncompact} and elsewhere, and their stability under variations
of the moduli \cite{stability, dcat, decat0, dstabmon, joyce2, joyce}
is of central importance.

In this paper I consider \slag\ submanifolds of noncompact \CY\ toric
varieties, and their restriction to a compact \CY\ embedded within it.
The linear sigma model construction and results of mirror symmetry are
used to study the properties of a certain class of these D-branes in
the two phases of the \Kah\ moduli space of the quintic hypersurface
in $\P^4$, as well as at special points in the complex structure
moduli space of the mirror manifold.

One of my aims in this paper is to show how existing results on
D-branes in conformal field theories may be obtained from the linear
sigma model picture which flows to (and interpolates between) these
conformal field theories in the infrared.  The general principle is
that quantities that are controlled under renormalization group flow
can be safely computed in the LSM framework: for example, since the
$(n,0)$-form $\Omega$ is holomorphic, its functional form is not
renormalized under RG flow, and we can compute string theory
quantities that depend on $\Omega$ within the linear sigma model.
Therefore A-type D-branes can be constructed in the LSM and descend to
A-type D-branes in string theory.  On the other hand quantities that
depend only on the \Kah\ structure are renormalized, and in general we
do not have direct control over or explicit knowledge of them in the
infrared limit of the LSM.

The main new results of this work are as follows:

At the Gepner point of \Kah\ moduli space, a class of \slag\ 
submanifolds of the linear sigma model target space -- those that span
the toric base of the target space, referred to as {\it base-filling
  D-branes} -- are shown to reproduce the labelling and intersection
properties of A-type rational boundary states of the Gepner model (the
boundary states first constructed by Recknagel and Schomerus \cite{rs}
and further developed in \cite{bstates}).  The corresponding set of
\slag\ submanifolds of the mirror reproduce the properties of the
B-type states at the Gepner point of the quintic, in accordance with
mirror symmetry (the analysis was performed for B-type states directly
in \cite{stringycy}, where they were associated to fractional branes
of the Landau-Ginzburg orbifold theory).

In this paper I will refer to the set of boundary states constructed
in \cite{rs, bstates} (which preserve the full tensor product \ntwo\ 
\SUSY\ algebra of the Gepner model) as ``rational boundary states'',
although it should be noted that there may be more general boundary
states of the Gepner models (which do not preserve the full algebra,
but only a diagonal \ntwo) which are also rational.

The Lagrangian D-branes of the Landau-Ginzburg model associated to a
single minimal model were obtained in \cite{dmirror} by studying BPS
solitons in Landau-Ginzburg theories; the toric geometry construction
naturally produces the extension of this result to the full Gepner
model.

The base-filling D-branes of the linear sigma model can be thought of
as providing a geometrical description of the rational boundary states
of the Gepner model, which are defined in abstract CFT and do not have
an obvious intrinsic geometrical description.  The construction of
rational boundary states following \cite{rs, bstates} relies on
preservation of the \ntwo\ \SUSY\ in each of the minimal model factors
of the Gepner model, and it appears difficult to construct a more
general class of boundary states within conformal field theory alone.
However, preservation of these extra symmetry algebras translates into
a simple geometrical property of the base-filling D-branes, and it
should be possible to relax this constraint to obtain a much larger
class of boundary states than have previously been constructed in CFT.

The base-filling D-branes have obvious relations between their
homology classes, and by representing the D-branes as polynomials
these relationships can be quantified in terms of relations between
the polynomials.  The large-volume homology classes of the B-type
rational boundary states of the quintic were computed in
\cite{quintic}, and the relations between these classes are reproduced
by the polynomial encoding.

The base-filling D-branes of the linear sigma model restrict to A-type
D-branes of string theory, and can be followed from the LG phase to
the geometrical phase.  They are shown to reproduce the construction
\cite{candelas, periods} of 3-cycles relevant for computing periods of
the compact \CY.

It was proposed in \cite{quintic} that some of the A-type states of
the quintic should be identified in the geometrical phase with certain
$\R\P^3$ submanifolds of the quintic, by comparing their intersection
forms.  The corresponding base-filling D-branes from the toric
geometry construction have the correct intersection properties, but do
not coincide with the $\R\P^3$s in the geometrical phase and instead
produce a distinct \slag\ submanifold with the same intersection form.
This is compatible with results from deformation theory.

The layout of the paper is as follows: section \ref{cft} contains a
brief review of some existing results on D-branes in conformal field
theories (rational boundary states in Gepner models, and
supersymmetric D-branes on \CYMs).  Section \ref{toric} provides an
introduction to toric geometry; the presentation follows that of
\cite{branestoric, holdiscs} and should be more accessible to
beginners than the usual mathematical treatments of toric geometry.
The linear sigma model \cite{wittenn2} is briefly recalled in section
\ref{lsm} as a natural consequence of the geometrical construction of
toric varieties.  The construction of A-type D-branes in the LSM is
discussed in section \ref{a-type}.  The properties of a certain class
of these D-branes are analyzed in the Landau-Ginzburg orbifold phase
of the LSM in section \ref{lg}, and in the non-linear sigma model
phase in section \ref{geombr}.  Section \ref{chargelatt} studies the
relationships that exist in the homology lattice of the D-branes.  The
possibilities for construction of new CFT boundary states based on the
geometry of the linear sigma model are discussed in section
\ref{newbranes}, and I bring together the results obtained in previous
sections with a proposal that the boundary states of the Gepner model
should be thought of as the ``latent geometry'' of the \slag\ 
submanifolds of the linear sigma model in a limit which confines the
theory to a single point.  Finally, section \ref{conc} summarizes some
unresolved problems and possibilities for further work.

\section{D-branes in Conformal Field Theories}
\label{cft}

There are two important classes of conformally-invariant string
compactification (conformal field theory): the non-linear sigma model
(NLSM), describing a string propagating on a Calabi-Yau manifold,
and the Gepner models, which are exactly solvable conformal field
theories built from a tensor product of \ntwo\ minimal models with the
correct central charge to saturate the conformal anomaly of string
theory.

The properties of D-branes in these two CFTs have been much studied in
recent years, and I will now review some of the relevant results.

\subsection{D-branes on Calabi-Yau manifolds}
\label{cym}

A BPS D-brane in Type II string theory preserves 1/2 of the spacetime
\SUSY; i.e.~it is invariant under an \nOne\ subalgebra of the \ntwo\ 
spacetime supersymmetry algebra.  The BPS conditions were worked out
from the point of view of both the worldsheet CFT and the target space
geometry in \cite{chernsimons, bbs, ooy}, which I now briefly review.

A Lagrangian submanifold $L$ of a \Kah\ manifold is one for which the
\Kah\ form pulls back to 0 on the submanifold:
\begin{equation}
\omega |_L = 0
\end{equation}
In terms of the worldsheet theory, D-branes that wrap Lagrangian
submanifolds preserve half of the worldsheet \SUSY\ (i.e.~they
preserve \ntwo\ worldsheet \SUSY), but not necessarily spacetime
\SUSY; the condition for preserving half of the spacetime
supersymmetry (i.e.~\nOne\ in spacetime) is that the submanifold must
further be {\it special} Lagrangian.

A \slag\ submanifold of a \CYM\ is a Lagrangian submanifold for which
the pullback of the holomorphic $n$-form of the \CY\ $n$-fold has a
constant phase on the submanifold:
\begin{eqnarray}
\hbox{Im log } \Omega|_L = \epsilon
\label{grade}
\end{eqnarray}
where $\epsilon$ is called the $U(1)$ grade of the D-brane and is
defined modulo $2 \pi$: it is associated to the relative phase of the
left- and right- spectral flow operators of the worldsheet theory when
they are glued together with an A-type automorphism on the boundary of
the worldsheet \cite{ooy}.  Two A-type D-branes that have the same
$U(1)$ grade will be mutually supersymmetric; D-branes with a
different grade will break spacetime supersymmetry and will therefore
not be stable (there exists a tachyon in the spectrum of open strings
stretching between the branes, which causes the system to decay into a
BPS system of D-branes with the same total topological charges,
e.g.~in the same total homology class).

B-type D-branes correspond to holomorphic submanifolds of the target
space, however this paper will not discuss B-type D-branes directly.
Under mirror symmetry the A- and B-type D-branes will interchange, so
the A-type states on $\cal M$ are exchanged with the B-type states on
the mirror manifold $\cal W$, and vice versa.  This operation also
interchanges the role of \Kah\ moduli and complex structure moduli on
the manifold and its mirror.  Therefore we can restrict our attention
to the A-type D-branes at the expense of considering both $\cal M$ and
deformations of one type (\Kah\ or complex structure) and $\cal W$
with deformations of the other type.  This is the approach I will take
in this paper.

A general construction of \slag\ submanifolds of a \CYM\ is not known,
but one known construction is the fixed-point set of a real involution
of the manifold (a ``reality condition'': the \slag\ submanifolds are
``real'' objects).

The example that will be used in section (\ref{geombr}) involves the
quintic hypersurface in $\P^4$ (a compact \CYM) defined by
\begin{equation}
\sum_{i=1}^5 z_i^5 = 0
\label{quintic}
\end{equation}
where $z_i$ are homogeneous coordinates on the $\P^4$.

This equation may be deformed by addition of monomials of degree 5;
these correspond to complex structure deformations of the \CYM.  The
complex structure deformation that is relevant for the mirror quintic
is:
\begin{equation}
\sum_{i=1}^5 z_i^5 -\psi z_1 \ldots z_5 = 0
\label{defquintic}
\end{equation}
where $\psi$ is a complex parameter.

If we impose the reality condition \cite{bbs}
\begin{eqnarray}
z_i &=& \overline{z_i} \nonumber \\
\Leftrightarrow \hbox{ Im } z_i &=& 0
\label{reality}
\end{eqnarray}
on each $z_i$, (i.e.~$z_i \equiv x_i \in \R$) then we obtain the real equation
\begin{equation}
\sum_{i=1}^5 x_i^5 = 0
\end{equation}
which has a unique solution for one of the coordinates in each
projective coordinate patch in terms of the remaining three.
Therefore, this 3-dimensional real submanifold is diffeomorphic to
$\R\P^3$.

More generally \cite{quintic} we can extend the reality condition
(\ref{reality}) to:
\begin{equation}
\hbox{Im } \omega_i z_i = 0
\label{reality2}
\end{equation}
where $\omega_i^5 = 1$, which gives a total of $5^{(5-1)} = 625$
$\R\P^3$s inside the quintic (since a common phase rotation $z_i
\mapsto \omega z_i$ acts trivially in projective space).

Another construction of \slag s when ${\cal M}$ is a toric variety
will be presented in section \ref{lg}.  There may be other real
involutions which can be imposed to construct \slag s, as well as more
general constructions.

\subsection{D-branes in Gepner Models}
\label{gepner}

The work of Recknagel and Schomerus \cite{rs} used the techniques of
Boundary Conformal Field Theory (BCFT) developed by Cardy
\cite{cardy1, cardy2}, Ishibashi \cite{ishibashi1, ishibashi2} and
others, to formulate states corresponding to D-branes in terms of
``boundary states'' of the world-sheet (open string) theory (in the
closed string channel).  Their construction was clarified and extended
in \cite{bstates} using the framework of simple current extensions.

If we choose to preserve the total tensor product algebra
${\cal{A}}^\otimes$, i.e.~the \ntwo\ superconformal algebra in each
minimal model factor of the Gepner model, the resulting CFT is
rational and its study is tractable.  Since all we require physically
is an overall \ntwo\ \SUSY, this construction preserves much more
symmetry than we need (it preserves a separate \ntwo\ algebra in each
minimal model factor), but preserving less symmetry renders the theory
non-rational and therefore difficult to study from the point of view
of CFT.

Rational boundary states of minimal models are in 1-1 correspondence
with the chiral primary fields of the bulk minimal model.  The Gepner
model boundary states constructed by Recknagel and Schomerus are
therefore labeled (before GSO projection) by
\begin{equation}
|\psi\rangle_\Omega = |L_1, \ldots, L_r; M_1, \ldots, M_r; S_1, \ldots, S_r\rangle_\Omega
\label{rslabel}
\end{equation}
where $\Omega=A, B$ labels which type of \SUSY\ is to be preserved by
the worldsheet boundary, $r$ is the number of minimal model factors of
the Gepner model, and
\begin{eqnarray}
L_i &\in& \{ 0, \ldots, k_i \} \nonumber \\ 
M_i &\in& \{ -k_i-1, \ldots, k_i+2 \} \nonumber \\ 
S_i &\in& \{ -1, 0, 1, 2 \nonumber \} \\
L_i + M_i + S_i &=& 0 \hbox{ mod } 2
\label{evengepner}
\end{eqnarray}
where $k_i$ is the choice of model in the \ntwo\ minimal series, for
the $i^{\rm th}$ tensor product factor.  The $S_i$ distinguish between
the NS ($S_i=0, 2$) and R sectors ($S_i=-1, 1$) of the minimal model,
and the two values in each sector correspond to a brane and its
anti-brane.

The labels (\ref{rslabel}) overcount the physically distinct boundary
states in several ways.  First, for a generic model there is a ``field
identification''
\begin{equation}
(L_i, M_i, S_i) \sim (k_i - L_i, M_i + k_i + 2, S_i + 2)
\label{fieldid}
\end{equation}
which reduces the number of distinct boundary states by half in each
minimal model factor\footnote{When one or more of the minimal model
  levels $k_i$ is even there are subtleties to do with fixed points
  under the field identification.  They have been studied in
  \cite{singcurv, bstates}, but I will not address these issues here
  since I mainly focus on the quintic model $(k=3)^5$.}.  This has a
simple geometrical interpretation in terms of the geometrical D-branes
derived later.

The choice of NS/R sectors are constrained by the GSO projection
(which ensures that the boundary states preserve \nOne\ spacetime
\SUSY) to be the same in each factor, and hence the distinct states
are labelled only by a single $S$ label \cite{quintic, bstates}.

There may be other over-counting depending on the symmetries of the
particular Gepner model chosen.  In general these will correspond to
geometrical symmetries of the linear sigma model that are inherited
by the conformal field theory at the infrared fixed point of each
phase (i.e.~which are unbroken by the vacuum submanifold of the linear
sigma model).

For example, the $(k=3)^5$ model corresponds to a point in the
extended \Kah\ moduli space of the quintic hypersurface in $\P^4$.
One finds that the A-type boundary states are physically
indistinguishable under a simultaneous shift $M_i \mapsto M_i + 2$ of
all of the 5 $M_i$ labels, and therefore this can be used to fix one
of the $M_i$ leaving four free.  This corresponds in the LSM to a
simultaneous $\Z_5$ rotation of each of the coordinates of the
D-brane, which acts trivially in the LSM target space as we will see
later.  In the LG orbifold phase of the quintic the target space is
$\C^5/\Z_5$ where the $\Z_5$ is precisely this common rotation, and in
the NLSM phase the target space is a line bundle over $\P^4$, and the
homogeneous coordinates on $\P^4$ are invariant under the same shift.

The B-type boundary states on the quintic are invariant under a
$\Z_5^4$ action on the $M_i$, which implies that the equivalence
classes are labeled by a single $M$ value.  Geometrically this is best
understood by looking at the symmetries of the mirror manifold, which
by the Greene-Plesser construction is given by an $(\Z_5)^3$ orbifold
of the quintic.  Therefore, the symmetry of the mirror models is
$(\Z_5)^4$ including the $\Z_5$ that acts ``projectively''.

The redundancies in the boundary state labeling will be explicit in
the geometrical representatives that will be constructed later.

The $U(1)$ grade of the boundary state $\theta$ is given by:

\begin{equation}
\frac{\theta}{\pi} \equiv \left( \sum_{i=1}^r (- \frac{M_i}{k_i+2} + \frac{S_i}{2}) - S_1 \frac{d}{4} \right) \mbox{ mod } 2
\label{gepnergrade}
\end{equation}

The intersection form of the D-brane boundary states can be computed;
in conformal field theory this is computed in the open string channel
by the Witten index
\begin{equation}
Tr_R(-1)^F
\end{equation}
for a worldsheet with two given boundary conditions (boundary states),
where $F$ is the worldsheet fermion number operator.

This quantity was computed for the rational boundary states in
\cite{quintic}.  For two such states
\begin{equation}
|L_1, \ldots, L_r; M_1, \ldots, M_r; S \rangle_\Omega \nonumber
\end{equation}
and
\begin{equation}
|\overline{L}_1, \ldots, \overline{L}_r; \overline{M}_1, \ldots, \overline{M}_r; \overline{S} \rangle_\Omega \nonumber
\end{equation}
the intersection form was found to be:
\begin{equation}
I_A = \frac{1}{C} (-1)^{\frac{S-{\overline S}}2} \sum_{\nu_0=0}^{K-1} \prod_{j=1}^r N_{{L_j}, \overline{L}_j}^{2 \nu_0 + M_j - \overline{M}_j}
\label{intera}
\end{equation}
for the A-type states, and
\begin{equation}
I_B = \frac{1}{C} (-1)^{\frac{S-{\overline S}}2} \sum_{m_{j'}} \delta^{(K')}_{\frac{M-M'}{2} + \sum \frac{K'}{2k_j + 4} (m_{j'} + 1)} \prod_{j=1}^r N^{m_{j'}-1}_{L_j, \overline{L}_j}
\label{interb}
\end{equation}
for the B-type states, where $C$ is a normalization constant,
and
\begin{eqnarray}
K' &=& \hbox{ lcm}\{k_j+2\} \nonumber \\
M &=& \sum_j \frac{K' M_j}{k_j + 2} \nonumber \\
\delta^{(K')}_{n} &\equiv& \delta_{(n \hbox{ mod } K')}
\end{eqnarray}
$N^{l}_{L_j, \overline{L}_j}$ are the extended $SU(2)_k$ fusion
coefficients \cite{quintic}:
\begin{eqnarray}
N^{l}_{L, \overline{L}} &=& \left\{ \begin{array}{ll}
      1 & \mbox{if $| L - \overline{L}| \leq l \leq $min$\{L+\overline{L}, 2k-L-\overline{L}\}$} \nonumber\\
      0 & \mbox{otherwise}
\end{array} \right. \nonumber\\
N^{-l-2}_{L, \overline{L}} &=& - N^{l}_{L, \overline{L}}
\label{fusion}
\end{eqnarray}

We will see later that these formulae can be derived easily from
geometrical considerations in the LSM.

\section{Toric Geometry}
\label{toric}

In this section I will give an intuitive description of the framework
of toric geometry using the ``symplectic quotient'' construction,
following \cite{branestoric} and \cite{holdiscs}.  This description
differs from the usual mathematical presentation of toric geometry,
but allows a more direct understanding of the geometrical
constructions of toric varieties and their (special) Lagrangian
submanifolds, as well as directly carrying over to the linear sigma
model construction: for any given toric variety expressed in this way,
the corresponding linear sigma model can be written down immediately.
See \cite{branestoric} for more details on the connections between
this framework and the more traditional approach to toric geometry.

Essentially, a toric variety is a $T^n$-fibration (hence the name {\it
  toric}) over some (not necessarily compact) linear base space with
boundary, where the $T^n$ fibres are allowed to degenerate over the
boundary of the base.  In the case where the base space is compact,
the resulting toric variety will also be compact.

A simple example of a toric variety is $\cn$, which can be parametrized
by $z_i = |z_i| e^{\imath \theta_i}, i = 1, \ldots n$.  This is a
\Kah\ manifold with \Kah\ form given by

\begin{eqnarray}
\omega &=& \imath \sum_i dz_i \wedge d{\bar z}_i \\
&=& \sum_i d( {|z_i|}^2) \wedge d\theta_i
\label{kahlercn}
\end{eqnarray}
A Lagrangian submanifold $L$ of a \Kah\ manifold $\M$ is defined by

\begin{equation}
\omega|_L = 0
\end{equation}
i.e.~the \Kah\ form vanishes identically upon restriction to the
submanifold.  Since we will be interested in Lagrangian (and \slag)
submanifolds in later applications to D-branes, it is convenient to
write $\cn$ in a way that makes the Lagrangian structure clear.

\begin{figure}[tbp]
\begin{center}
\epsfig{file=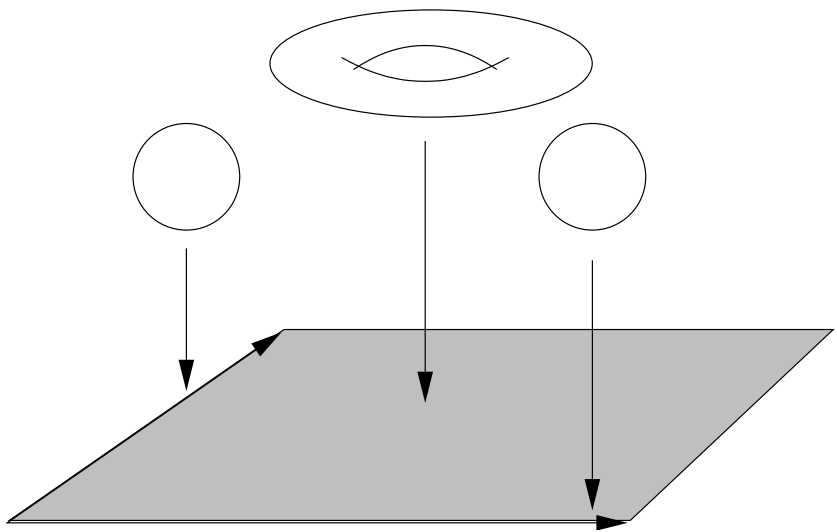}
\parbox{5.5in}{
\caption{\small{$\C^2$ as a $T^2$ fibration over $(\R^+)^2$, where the fibre degenerates over the boundary of $(\R^+)^2$, as one or more of the $S^1$ shrink to zero size.\label{cntoric}}}
}
\end{center}
\end{figure}

Using the natural U(1) action
\begin{equation}
z_i \mapsto z_i e^{\imath \theta_i}
\end{equation}
we can express $\cn$ as a $T^n$ fibration over a Lagrangian
submanifold $L$ of $\cn$ defined by taking $\theta_i$ constant for all
$i$; $L$ is then isomorphic to the positive segment of $\rn$
parametrized by $|z_i|^2\geq 0, i = 1, \ldots, n$ (see figure
\ref{cntoric}).  We parametrize the base by $|z_i|^2$ instead of
$|z_i|$ in order to make the base space linear\footnote{The projection
  to the base space $z_i \mapsto |z_i|^2$ is also called the {\it
    moment map.}}.
The $T^n \simeq U(1)^n$ action acts on the $|z_i|^2$ to recover the
manifold $\cn$.  The boundary of $L$ is given by the union of the
hyperplanes $|z_i|^2 = 0$, and the U(1) action has fixed points along
each of these boundary segments, corresponding to the degeneration of
the $T^n$ fibre to a $T^{n-d}$ at points on the boundary where $d$
hyperplanes intersect.

Note that in choosing this parametrization of $\cn$ and fixing each of
the angular coordinates $\theta_i$, the \Kah\ form (\ref{kahlercn}) in
fact vanishes term by term, since $d\theta_i \equiv 0$ for all $i$.

In order to obtain a more general toric variety of complex dimension
$n$ we use the method of {\it symplectic quotient} $\C^{n+r} // G$,
where $G \simeq U(1)^r$ for some $r$.  Concretely, we proceed in two
steps: first restricting to a certain linear subspace in the $|z_i|^2$
and then dividing by the group action $G$.

The choice of subspace and group action $G$ is defined by a set of $r$
vectors of integral weights or ``charges''\footnote{The term comes
  from the related linear sigma model construction to be described in
  section \ref{lsm}, where the $Q^a$ define the charges of the
  LSM chiral superfields under the $U(1)^r$ gauge group.}
\begin{equation}
Q^a = (Q_1^a, \ldots, Q_{n+r}^a),\ a = 1, \ldots, r
\end{equation}
For each $a$ we define the hyperplane
\begin{equation}
\sum_{i=1}^{n+r} Q_i^a |z_i|^2 = r^a
\label{dterm}
\end{equation}
The intersection of these $r$ hyperplanes is (assuming the $Q^a$ are
linearly independent vectors) a real space $D$ (with boundary) of
dimension $n+r-r = n$.  Each parameter $r^a$ is a deformation modulus
for the toric variety (it is just the normal translation modulus for
the hyperplane).

In order to obtain a \Kah\ manifold we quotient the $T^{n+r}$ bundle
over $D$ by the $U(1)^r$ action generated by a simultaneous phase
rotation of the coordinates:
\begin{equation}
z_i \mapsto e^{\imath Q_i^a \epsilon^a} z_i
\label{quotientu1}
\end{equation}
for each $a = 1, \ldots, r$, where $\epsilon^a$ are the
generators of the $U(1)$ factors.  This fixes $r$ of the phases and
gives a $T^{n}$ bundle over $D$.  This construction preserves the
\Kah\ form on the toric variety induced from (\ref{kahlercn}), and the
quotient space is therefore a \Kah\ manifold of complex dimension $n$.
The translation moduli $r^a$ become the \Kah\ moduli of the \Kah\ 
manifold, so the manifold has $\hbox{dim}_\C H^{1,1}(M) = r$.  When
$r=1$ we obtain a weighted projective space with weights $Q_i$ (if one
or more of the $Q_i$ are negative then this is a non-compact
generalization of the usual compact weighted projective spaces); the
cases $r > 1$ are more general toric varieties.

If in addition the charges $Q^a$ satisfy
\begin{equation}
\sum_{i=1}^{n+r} Q_i^a = 0
\label{cycharges}
\end{equation}
for all $a$, then the toric variety is moreover a \CYM, since the
holomorphic $(n+r,0)$-form of $\C^{n+r}$ is $\C^*$-invariant and pulls
back to a non-vanishing $(n,0)$-form on the toric variety
\cite{wittenn2}.  Note that (\ref{cycharges}) implies that for each
$a$ at least one of the $Q_i^a$ must be negative: this
implies that the \CYM\ will therefore be non-compact since the
hyperplane (\ref{dterm}) is unbounded above in this coordinate
direction along $D$.

In later sections we will make use of the non-compact \CY\ $n$-fold
\begin{equation}
{{\cal O}_{\P^{n-1}}}(-n)
\end{equation}
the holomorphic line bundle of degree $-n$ over $\P^{n-1}$
(equivalently, the normal bundle to $\P^{n-1}$). Starting with
$\C^{n+1}$ as a $T^{n+1}$ fibration over $L = (\R^+)^{n+1}$ we choose
a real codimension-1 subspace of the base space $L$ defined by the
vector $Q = (1, 1, \ldots, 1, -n)$, i.e.~the hyperplane

\begin{equation}
\sum_{i = 1}^{n+1} |z_i|^2 - n |z_{n+1}|^2 = r
\label{hyperplane}
\end{equation}
where $r$ is an arbitrary real parameter.

\begin{figure}[tbp]
\begin{center}
\epsfig{file=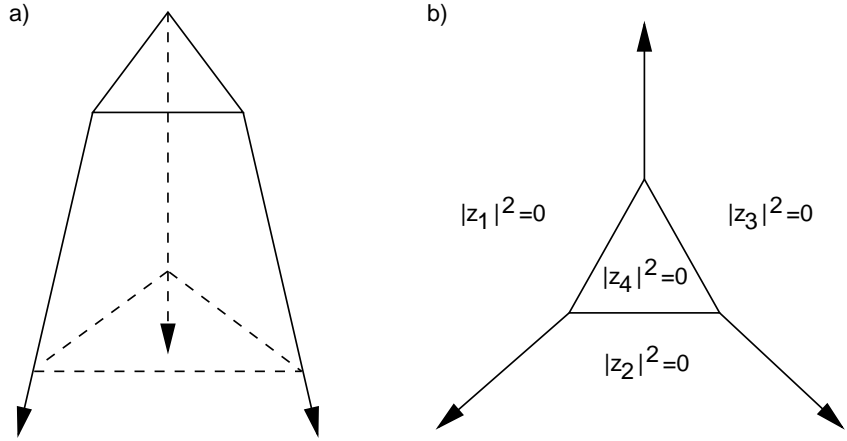}
\parbox{5.5in}{
\caption{\small{a) The geometry of the base of ${\cal{O}}_{\P^2}(-3)$ as a subset of $\R^3$. b) The same geometry projected onto the plane, showing the 2-dimensional boundary of the base space.  The plane is divided into various regions labeled by the coordinates that vanish in each, describing the embedding of the plane into the various boundary hyperplanes of $(\R^+)^4$.\label{p2}}}
}
\end{center}
\end{figure}

At $|z_{n+1}|^2=0$ the equation becomes
\begin{equation}
\sum_{i = 1}^{n} |z_i|^2 = r
\end{equation}
which defines an $(n-1)$-simplex of size $r$ (see Figure \ref{simplex}).
The size of the simplex increases along the $|z_{n+1}|^2$ direction.
Therefore, this subspace describes a family of $(n-1)$-simplices
parametrized by $|z_{n+1}|^2$:
\begin{figure}[tbp]
\begin{center}
\epsfig{file=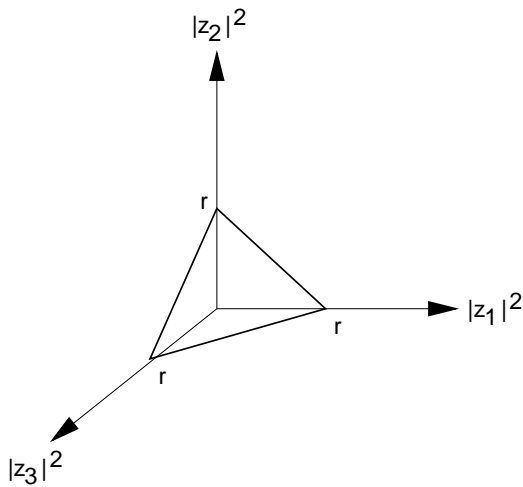}
\parbox{5.5in}{
\caption{\small{The simplex at $|z_4|^2=0$ for the case $n=3$.  For nonzero $|z_4|^2$ the simplex has size $r+|z_4|^2$\label{simplex}}}
}
\end{center}
\end{figure}
\begin{equation}
\sum_{i = 1}^{n} |z_i|^2 = r + n |z_{n+1}|^2
\label{linedterm}
\end{equation}
as shown in Figure \ref{p2}.  The base space therefore has one
non-compact direction and $n-1$ compact directions.

After dividing by the $U(1)$ symmetry, we are left with a
$T^n$-fibration over a real $n$-dimensional base space, which is a
toric \CYM.  As before, the $T^n$ fibration will degenerate along the
boundary of the base.  The total space over the $(n-1)$-simplex at
$z_{n+1}=0$ is isomorphic to $\P^{n-1}$ by construction: along this
face we have $z_{n+1} = 0$, and the symplectic quotient construction
fixes the radius $|z_1|^2 + |z_2|^2 + \ldots + |z_n|^2 = r$ of an
$S^{2n-1}$ inside $\C^n \simeq \R^{2n}$, as well as a $U(1)$ acting by
common phase rotation.  Altogether we have taken the quotient of
$\C^n$ by a $\C^*$ action; this is the usual construction of
$\P^{n-1}$.

The total space of the non-compact direction transverse to the simplex
is given by an $S^1$ fibre over $\R^+$.  Therefore, corresponding to
every point in $\P^{n-1}$ we have a copy of $\C \simeq \R^+ \times
S^1$.  The toric variety is therefore a (holomorphic) line bundle, and
the $U(1)$ action on the coordinate $z_{n+1}$ defined by the charge
$Q_{n+1} = -n$ means it is the promised line bundle ${{\cal
    O}_{\P^{n-1}}}(-n)$.  Alternatively, the space can be viewed as
the weighted projective space $\WP^n(1, 1, \ldots, 1, -n)$ (which
contains $\P^{n-1}$ as a compact submanifold).

\begin{figure}[tbp]
\begin{center}
\epsfig{file=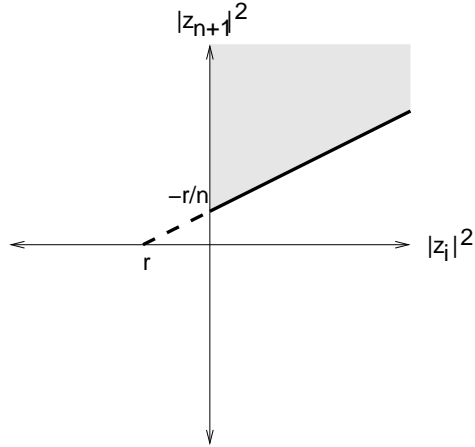}
\parbox{5.5in}{
\caption{Cross-section of the toric base of the $Q=(1,1,1,\ldots, 1, -n)$ toric variety showing how the target space geometry changes for $r \leq 0$.  At $r=0$ the simplex at the ``tip'' of the toric base (shown in figures \ref{simplex} and \ref{p2}) shrinks to zero size, and the topology of the space changes from ${{\cal O}_{\P^{n-1}}}(-n)$ to $\C^n/\Z_n$.  Throughout the phase $r \leq 0$ the target space geometry stays the same up to a shift along the $z_{n+1}$ axis, because of the requirement that $|z_i|^2 \geq 0$.\label{lgfig}}}
\end{center}
\end{figure}

As we take $r \rightarrow 0$, the simplex forming the base of the
$\P^{n-1}$ shrinks to zero size, and for $r < 0$ the geometry is
isomorphic, up to a translation along the $|z_{n+1}|^2$ axis (see
Figure \ref{lgfig}).  Geometrically, taking $r \rightarrow 0$
corresponds to ``blowing down'' the $\P^{n-1}$ at the base space of
the line bundle $\linen$, and the geometry becomes isomorphic to
$\C^{n}/\Z_{n}$.  This transition changes the topology of the space,
and is an example of a {\it birational equivalence} \cite{wittenn2}.
The induced metric from $\C^{n+1}$ on the line bundle is not
Ricci-flat; however in this case the Ricci-flat metric is known
explicitly and is given by the Calabi metric \cite{newmanifolds}
\begin{equation}
ds^2 = {\frac{1}{\rho}} \left\{ dy_i dy^i - {\frac{1}{\rho}} y_i dy^i y_{\bar j} dy^{\bar j} \right\}
 + \left\{ \rho^n dw d{\bar w} + n \rho^{n-1}\left({\bar w} dw y_{\bar i} dy^{\bar i} + w d{\bar w} y_i dy^i\right) \right\}
\label{calabimet}
\end{equation}
where $\rho = 1 + y_i y^i$.  The coordinate $w$ parametrizes the
non-compact direction of the line bundle, and the metric
(\ref{calabimet}) reduces at $w=0$ to the Fubini-Study metric on
$\P^{n-1}$:
\begin{equation}
ds^2 = {\frac{1}{\rho}} \left\{ dy_i dy^i - {\frac{1}{\rho}} y_i dy^i y_{\bar j} dy^{\bar j} \right\}
\label{fubinistudy}
\end{equation}
where $y_i,\ i = 1, \ldots, n-1$ are inhomogeneous coordinates on the
$\P^{n-1}$, in a coordinate patch ${\cal U}_n$.

The linear sigma model for the line bundle, which I discuss in the
next section, will have the induced non-Ricci-flat target space metric
at high energies, but it is expected to flow to the Calabi metric
under worldsheet RG flow to the infrared, i.e.~the linear sigma model
becomes conformal in this limit (and correctly describes perturbative
string theory on the line bundle).

For a given non-compact \CYM\ described by a line bundle over a
$\P^{n-1}$ (or $\WP^{n-1}$) we can obtain a family of compact \CYMs\ 
of complex dimension $n-2$ by restricting to the zero locus of a
quasi-homogeneous degree-$n$ polynomial inside the $\P^{n-1}$ (more
generally, we can consider ``complete intersections'' of multiple
polynomials).  From the point of view of the compact \CY, the presence
of the line bundle is irrelevant, but it is needed for the LSM
construction.  The simplest case for $\P^{n-1}$ is a single degree-$n$
polynomial of Fermat type:
\begin{equation}
\sum_{i=1}^{n} {z_i}^n = 0
\end{equation}
which is a \CY\ hypersurface of complex dimension $n-2$ within the
$\P^{n-1}$.
 
Note that in general these hypersurfaces do not respect the Lagrangian
or toric descriptions of the ambient space in which they are embedded.
However a toric description may be recovered in a limit when the
hypersurface degenerates , such as $\psi \rightarrow \infty$ for the
mirror quintic (\ref{defquintic}) \cite{branestoric}.  In this limit
the defining equation of the quintic (\ref{defquintic}) becomes (after
scaling out by $\psi$)
\begin{equation}
z_1 \ldots z_r = 0
\end{equation}
which is solved by taking one or more of the $z_i = 0$.  In terms of
the toric description of $\P^4$ the quintic is restricted to the
boundary of the 4-simplex, and the $T^3$ fibration over this boundary
produces (before dividing out by the extra orbifold symmetry) 5
intersecting $\P^3$s (faces of the 3-skeleton of the 4-simplex),
which intersect in 10 $\P^2$s (faces of the 2-skeleton), which in turn
meet in 10 $\P^1$s (faces of the 1-skeleton), which meet in 5 points
(faces of the 0-skeleton) \cite{candelas}.

Just as in the non-compact case, the metric on a hypersurface
inherited from the ambient $\P^{n-1}$ is not Ricci-flat, even if we
use the Ricci-flat metric on the $\linen$ in which $\P^{n-1}$ is
embedded.  Unfortunately, in contrast to the non-compact case the
Ricci-flat metric is not known explicitly for any compact \CYMs\ of
dimension 3 or higher (although its existence is guaranteed by Yau's
theorem).  However, the holomorphic $n$-form of the non-compact
ambient space (which is obtained by pullback from the original
$\C^{n+1}$) pulls back to the correct holomorphic $(n-2)$-form on the
hypersurface.

In order to later check whether a submanifold is special Lagrangian
one needs the explicit form of the holomorphic $n$-form on the \CY.
This is induced from the holomorphic $(n+r)$-form on $\C^{n+r}$
\begin{equation}
\Omega^{(n+r)} = dz_1 \wedge \ldots \wedge dz_{n+r}
\end{equation}
For a general weighted projective $n$-space $\WP^{n}(k_1, \ldots,
k_n)$ obtained by projection from $\C^{n+1}$, an $n$-form is obtained
\cite{complexman} by contraction of $\Omega^{(n+1)}$ with the vector
field that generates the $\C^*$ action
\begin{equation}
(z_1, \ldots, z_{n+1}) \sim (\lambda^{k_1} z_1, \ldots, \lambda^{k_{n+1}} z_{n+1})
\label{cstar}
\end{equation}
giving
\begin{equation}
\Omega^{(n)} = {\frac{1}{{(N+1)!}}} \epsilon^{i_1 \ldots i_{n+1}} k_{i_{n+1}} z_{i_{n+1}} dz_{i_1} \ldots dz_{i_n}
\label{omega}
\end{equation}
For a general projective space $\Omega^{(n)}$ is not well-defined,
since under the $\C^*$ action (\ref{cstar}) it transforms like
\begin{equation}
\Omega^{n} \mapsto \lambda^{\sum_{i=1}^{n+1} k_i} \Omega^{n}
\label{scalingomega}
\end{equation}
In the special case where $\sum_{i=1}^{n+1} k_i = 0$ it is globally
well-defined; these are precisely the line bundles described above.
$\Omega^{(n)}$ can further be shown to be non-vanishing; therefore
these spaces are non-compact \CYMs.

In order to produce a well-defined $n$-form for a compact $\WP^n$ we
can choose a collection of polynomials $P_1, \ldots, P_\alpha$ and we
take instead
\begin{equation}
{\Omega^{(n-\alpha)}}' = \int_\Gamma {\frac{\Omega^{(n)}}{P_1 \ldots P_\alpha}}
\end{equation}
where $\Gamma$ is a real $\alpha$-dimensional contour that is the
product of small circles around each of the surfaces $P_i = 0$, so the
integral picks up the residue coming from the poles at $P_i=0$.  This
form will be scale-invariant, and hence globally defined, if the $P_i$
are chosen to have appropriate degree to compensate for the
transformation of $\Omega^{(n)}$.  It can be shown that the algebraic
variety defined by the intersection of the polynomial vanishing loci
$\{P_i = 0\}$ is a compact \CYM.

\section{Linear Sigma Models}
\label{lsm}

The linear sigma model was introduced in \cite{wittenn2}, and will not
be described in detail here.  The reader will observe that the
construction in \cite{wittenn2} precisely follows the symplectic
quotient construction of non-compact toric \CYMs\ from the previous
section: a toric \CY\ variety is obtained as the vacuum manifold of
the theory (parametrized by the scalar fields), with the linear,
i.e.~``wrong'' metric on the \CY\ in the UV; localization to a compact
\CY\ hypersurface within the non-compact \CY\ is implemented by an
appropriate superpotential term in the LSM.  Therefore using the
construction of the previous section we can use the associated linear
sigma model to describe string propagation on non-compact toric \CY\ 
$n$-folds\footnote{Studying non-compact manifolds is useful for
  providing local descriptions of string compactifications where we
  neglect the rest of the compact manifold ``at infinity'', e.g.~for
  studying the neighbourhood of a singularity.}, as well as compact
\CYMs\ of lower dimension that can be embedded in a non-compact \CY\ 
as a hypersurface or complete intersection of hypersurfaces.

Essentially, under RG flow to the infrared the target space metric of
the $d=2$ linear sigma model flows from the linear metric to the
Ricci-flat metric (and the coupling constant tends to infinity,
localizing onto classical vacua of the theory).  We can avoid the
complexities of working with the Ricci-flat metric directly by using
the linear sigma model, providing we only work with quantities that
are protected or controlled under RG flow, so that we can follow them
to the CFT limit.  For example, since the $(n,0)$-form $\Omega$ is
holomorphic, its functional form is not renormalized under RG flow,
and we can therefore hope to identify \slag\ submanifolds of the
linear sigma model target space with A-type boundary states of the
conformal field theory upon restriction to the vacuum submanifold of
the LSM.

The real \Kah\ moduli $r$ of a toric variety are complexified by the
$\theta$-angles of the LSM (which becomes the B-field in string
theory) through the combination ${\frac{\theta}{{2 \pi}}} + \imath r$,
and for the 1-parameter models the complexified \Kah\ moduli space has
two phases.  When $r>0$ the infrared fixed point of the linear sigma
model is a non-linear sigma model with the ``correct'' Ricci-flat
metric on the target space (e.g.~$\linefour$, or the quintic
hypersurface inside the $\P^4$) and this is called a geometrical
phase.  The phase $r<0$ corresponds formally to an analytic
continuation to negative \Kah\ class.  For $\linen$ this means
``negative size'' of the $\P^{n-1}$ in which the \CY\ hypersurface is
embedded, i.e.~we pass to the blown-down phase where the $\P^{n-1}$
has been collapsed to a point, and the target space is $\C^n/\Z_n$
(the \Kah\ modulus for moving around in this phase is hidden within
the kinetic term of the LG model, which is only defined implicitly
through RG flow).  If we also have a superpotential in the theory then
this phase is a Landau-Ginzburg orbifold theory.  The singularity at
$r=0$ can be avoided by turning on a non-zero $\theta$-angle.

The Gepner model exists at the infrared fixed point of the LSM in the
limit $r \rightarrow -\infty$, the ``deep interior point'' of the LG
phase of \Kah\ moduli space, and is an exactly solvable CFT.  In the
opposite limit $r \rightarrow \infty$ of the geometrical phase (the
``large volume limit''), closed string instanton corrections are
suppressed since the volume of all 2-cycles is large, and the infrared
fixed point is a nonlinear sigma model described by classical
geometry.

A-type D-branes may decay under variation of complex structure
\cite{joyce}, but are stable under \Kah\ deformations \cite{quintic,
  stability}.  Therefore when we consider A-type D-branes in the
various phases of \Kah\ moduli space, they will remain stable
throughout since the special Lagrangian condition that defines A-type
branes depends on the complex structure, which is kept fixed.

Mirror symmetry of \CY\ manifolds exchanges A- and B-type D-branes, as
well as the role of \Kah\ and complex structure moduli on the manifold
and its mirror.  Therefore, in order to understand the behaviour of
B-type D-branes under variation of \Kah\ structure we can equivalently
consider A-type D-branes of the mirror theory under variation of
complex structure.  There are issues of stability of these A-type
D-branes to consider, but this is a purely geometrical problem (in
contrast to the B-type D-branes, which are destabilized by instanton
effects).  It is possible to obtain concrete results in certain
limits, for example the mirror to the deep interior points of the
\Kah\ moduli space of the quintic.

Mirror symmetry can be studied in the LSM framework, and essentially
has the interpretation of T-duality along the $T^n$ fibres of the
non-compact toric variety \cite{mirrorlsm}.  Starting with the LSM for
a given \CYM\ (including compact \CYMs\ embedded in a non-compact \CY\ 
via a superpotential), the authors of that paper were able to derive
the corresponding LSM for the proposed mirror \CY\ (as well as the
mirror for more general situations).  I will not make use of their
derivation explicitly, but will instead use known results on mirror
symmetry for the quintic (see \cite{greeneplesser} \cite{candelas}).

\subsection{A-type D-branes in the LSM}
\label{a-type}

Recall from section \ref{cym} that A-type D-branes are associated to
\slag\ $n$-cycles of a \CY\ $n$-fold.  We would like to realise a
class of A-type D-branes in the linear sigma model; using the
symplectic quotient construction this is equivalent to describing
\slag\ submanifolds of the toric variety.  The starting point of the
symplectic quotient construction was the description of $(\R^+)^n$ as
a Lagrangian submanifold of $\C^n$: we can immediately describe other
Lagrangian and \slag\ D-branes as submanifolds of this space (there
may be other possibilities that cannot be obtained by this method).

Lagrangian submanifolds are obtained by taking additional hyperplane
constraints in the toric base \cite{holdiscs}: the intersection of $p$
linearly independent hyperplanes will give an $(n-p)$-dimensional
subspace of the base, and taking the orthogonal subspace of the fibres
gives a $T^{p}$ fibration over this base space, producing a real
$n$-dimensional submanifold of the complex $n$-dimensional toric
variety.

Each hyperplane (indexed by $\alpha$) is defined by a normal vector
${\vec{q}}\,^\alpha$ and a translation modulus $c^\alpha$ fixing the
location of the hyperplane:
\begin{equation}
\sum_{i=1}^{n+r} q_i^\alpha |z_i|^2 = c^\alpha,\ \alpha = 1, \ldots, p
\label{slaghyper}
\end{equation}
To obtain a rational (Hausdorff) subspace of the $T^n$ the entries
$q_i^\alpha$ of the normal vector are constrained to be integers.
These Lagrangian submanifolds are therefore characterized by an
integer $p$ which specifies the number of ``D-term-like'' constraints
defining the base of the submanifold as an intersection with the base
of the toric variety.

The orthogonality conditions on the angular coordinates which define
the $T^{p}$ fibre of the Lagrangian submanifold are \cite{holdiscs}
\begin{equation}
\vec{v}\,^\beta \cdot \vec{\theta} = 0 \hbox{ mod } 2 \pi,\ \beta=1, \ldots, n-p
\label{fibre}
\end{equation}
where $\vec{\theta} = (\theta_1, \ldots, \theta_{n+r})$ are the
angular coordinates on $T^{n+r}$, and $\vec{v}\,^\beta$ are integral
vectors that span the intersection of the hyperplanes, i.e.~which
satisfy
\begin{equation}
\vec{v}\,^\beta \cdot \vec{q}\,^\alpha = 0
\end{equation}
In order to be well-defined after dividing by the $U(1)^r$
gauge-symmetry of the toric variety we also require the $\vec{v}\,^\beta$ to
satisfy
\begin{equation}
\vec{v}\,^\beta \cdot \vec{Q}\,^a = 0
\label{normalcy}
\end{equation}

The condition (\ref{cycharges}) for a toric variety to be \CY\ is
\begin{equation}
\sum_{i=1}^{n+r} Q_i^a = 0,\ a = 1, \ldots, r
\label{cycond}
\end{equation}
which has a similar form to the constraint that a Lagrangian
submanifold be \slag\footnote{This formal similarity seems to be at
  the foundation of recent studies of open string mirror symmetry
  (mirror symmetry for \CYMs\ including D-branes) \cite{openclosed,
    openmirror}, in which a non-compact \CY\ 3-fold together with a
  certain type of \slag\ D-brane ($p=2$ in my notation) is promoted
  into a \CY\ 4-fold without D-branes, to which closed string mirror
  symmetry can be applied to compute exact disc instanton sums of the
  original theory.  This is called ``open/closed string duality'' in
  \cite{openclosed}.}:
\begin{equation}
\sum_{i=1}^{n+r} q_i^\alpha = 0,\ \alpha = 1, \ldots, p
\label{slagcond}
\end{equation}

In this paper I will focus on the class $p=0$: in the next section
these submanifolds will be related to the set of rational boundary
states of the Gepner model.  They are submanifolds with no additional
hyperplane constraints in the base and which will therefore span the
toric base of the toric variety: I will sometimes refer to them as
``base-filling D-branes'' to emphasize this property.  If the toric
variety is \CY\ then these submanifolds are furthermore \slag\ 
submanifolds.  To obtain the $p=0$ submanifolds we can choose $n$
vectors $\vec{v}\,^\beta$ that span the hyperplane defining the \CY,
i.e.~which satisfy (\ref{normalcy}).

Recall that $\linen$ is described by a single set of charges
\begin{equation}
Q = (1, 1, \ldots, 1, -n)\ ,
\end{equation}
which gives one D-term constraint (which fixes the base of the
$T^{n+1}$ fibration to lie within a hyperplane in $(\R^+)^{n+1}$), and
one $U(1)$ gauge invariance (which reduces the fibre from $T^{n+1}$ to
$T^n$).  If we are interested in studying D-branes on a compact \CY\ 
such as the quintic hypersurface in $\P^4$, we need to consider \slag\ 
submanifolds of $\linen$ that intersect the $\P^{n-1}$, as well as the
hypersurface within it.  Submanifolds that do not intersect the
hypersurface will not be visible to the string theory at the infrared
fixed point, which is constrained to lie within the hypersurface.  In
the LG phase the constraint is that the submanifolds must intersect
the orbifold point.

For $\linen$ we can take the $\vec{v}\,^\beta$ to be
\begin{eqnarray}
\vec{v}\,^1 &=& (n, 0, \ldots, 0, 1) \nonumber \\
\vec{v}\,^2 &=& (0, n, \ldots, 0, 1) \nonumber \\
&\vdots& \nonumber \\
\vec{v}\,^n &=& (0, 0, \ldots, n, 1)
\end{eqnarray}
These span the hyperplane (although they do not form an orthonormal
basis).  The fibre constraints (\ref{fibre}) reduce to
\begin{equation}
n \theta_i + \theta_{n+1} = 2 \pi a_i,\ a_i \in \Z,\ i = 1, \ldots, n
\end{equation}
or
\begin{equation}
\theta_i = {\frac{2 \pi a_i - \theta_{n+1}}{n}}
\end{equation}
Using the $U(1)$ symmetry (\ref{quotientu1}) we can set $\theta_{n+1}
= 0$; the $\theta_i$ become:
\begin{equation}
\theta_i = {\frac{2 \pi a_i}{n}}
\end{equation}
i.e.~they are $n^{\rm th}$ roots of unity.  These are the same
constraints obtained in \cite{dmirror} for D-branes of the
Landau-Ginzburg theory associated to a single \ntwo\ minimal model
(i.e.~single $\theta$), which were obtained by studying the BPS
solitons of the LG theory.  Since the Gepner model is constructed from
the tensor product of \ntwo\ minimal models (with certain
identifications and projections) it is natural to expect that the
rational boundary states should have a similar form in the tensor
product theory, however note that they are derived here from pure
geometry.

Along the submanifold the values of $\theta_i$ are constant.
Therefore, these submanifolds are isomorphic to the toric base of the
\CY, and in particular they have a boundary isomorphic to the boundary
of the base.  For comparison to existing results in the literature to
which I will later relate them, I will refer to these $n$-dimensional
submanifolds with boundary (i.e.~submanifolds which are described by
an $n^{\rm th}$ root of unity in each angular coordinate) as
``spokes''.  The presence of the boundary means that these
submanifolds are 3-chains, and one needs to take appropriate
differences of two such branes in order to obtain a \slag\ 3-cycle,
and I will therefore also refer to them as ``half-branes'' as context
suggests.

Since this class of \slag\ submanifold spans the simplex that is the
toric base of the $\P^{n-1}$, appropriately chosen \slag s will
intersect the hypersurface embedded within it.  However in order to
obtain a good D-brane, we need to take the difference of two such
submanifolds with a common boundary in the vacuum submanifold of the
LSM, so that a string in the infrared will see a D-brane with no
boundary.  For this D-brane to be BPS the two half-branes must have
the same $U(1)$ grade (\ref{grade}); i.e., for A-type D-branes the
pair of submanifolds with common boundary must also be a \slag\ 
submanifold.

As will be shown in the next section, this class of A-type D-branes in
the LSM is in 1-1 correspondence with the set of A-type boundary
states of the Gepner model constructed in \cite{rs, bstates} and
reproduces their symmetries and intersection form.  When we blow up
the origin of $\C^n/\Z_n$ (i.e.~pass to the NLSM phase) and restrict
to the \CY\ hypersurface in the blown up $\P^{n-1}$ it is possible to
relate a particular subset of the blown-up LSM D-branes to known
\slag\ cycles in the compact \CY\ hypersurface.

By construction, a pair of half-branes will have a constant phase of
the holomorphic $n$-form for each half-brane, but in general it need
not be the same phase for both.  In other words, the pair of
submanifolds are only piecewise \slag, not necessarily globally \slag.
From the point of view of the worldsheet theory, since the piecewise
\slag\ submanifolds are only Lagrangian submanifolds they only preserve
\ntwo\ world-sheet \SUSY, and do not preserve space-time \SUSY.

Since they are not BPS objects, we expect there to be a tachyon in the
string excitation spectrum of a single brane which will drive a flow
to a state in the same homology class that is BPS and therefore stable
(tachyon-free).  In other words, there should be a \slag\ submanifold
in the same homology class as the non-\slag\ submanifold we started
with.  {\it A priori} this may not be a \slag\ submanifold of the type
considered here (i.e.~a pair of spokes aligned along $n^{\rm th}$
roots of unity), but in fact for each homology class obtained by the
spoke construction for the quintic there exists a \slag\ submanifold
in the same homology class that is also a pair of spokes.  Therefore,
even the pairs of spokes that are not themselves \slag\ are
representatives of other spoke pairs in the same homology class that
are \slag.

The A-type D-brane construction described here is only strictly valid
at the ``deep interior point'' of each LSM phase (the large volume
limit for the geometrical phase, and the Gepner point for the LG
phase): away from the large volume limit the superpotential of the
D-brane world-volume theory receives instanton corrections coming from
open string worldsheets ending on the D-brane (which wind a nontrivial
$S^1$ of the D-brane and wrap a nontrivial Riemann surface in the
bulk), and control is also lost away from the Gepner point of the LG
phase.  The disc instanton corrections have been analyzed for
non-compact \CYMs\ in \cite{openinst, mirroropen, holdiscs,
  openclosed, openmirror, dualityweb, disclsm} using open string
mirror symmetry, for the \slag s\ labeled by $p=2$ in the notation
used above, but this analysis has not yet been extended to the other
$p$ classes.  Contributions to the instanton-generated superpotential
for the quintic were analyzed in the B-model in \cite{quinticW}.

\subsection{D-branes in the LG phase}
\label{lg}

In this section I specialize to the quintic for definiteness, although
the results should generalize in a straightforward manner to the other
\CYMs\ with $h^{1,1}=1$, except as noted.  In the Landau-Ginzburg
orbifold phase of the quintic, we will consider pairs of half-branes
of the type constructed in the previous section, and I will show how
they are related to the A-type rational boundary states at the Gepner
point.  The corresponding construction of B-type boundary states will
be related to A-type states in the mirror Landau-Ginzburg theory
\cite{greeneplesser}, in accordance with mirror symmetry.

In the LG phase the target space is $\C^5/\Z_5$, and the \slag\ 
submanifolds (half-branes) constructed in the previous section are
particularly simple.  In each coordinate $z_i$ of the orbifold space
$\C^5/\Z_5$ they are parametrized along rays that are aligned along
fifth roots of unity through the origin (modulo the $\Z_5$ orbifold
symmetry that acts by a common phase rotation of the $z_i$).  The rays
are oriented, inducing an orientation for the submanifold.  Together,
the independent coordinates $r_i$ along the 5 rays parametrize an
oriented 5-dimensional real submanifold of the target space, which is
isomorphic to the base of $\C^5/\Z_5$ in the toric construction of
section \ref{toric}, and has a boundary over the boundary of the toric
base space (when one or more of the coordinates $r_i = 0$).  The
submanifolds are referred to as ``spokes'' because their image in each
of the coordinates $z_i$ is a 1-dimensional ray, but note that the
submanifolds as a whole are 5-dimensional objects.

The $\theta_i$ are angular coordinates on $\C^5$, i.e.~they are not
single-valued on $\C^5/\Z_5$.  However, when we fixed $\theta_6 = 0$
there was a residual $\Z_5$ symmetry left unfixed, which acts on the
$\theta_i$ by a common $\Z_5$ phase rotation.  Acting with this
symmetry to bring $\theta_1$ into the range $[0, {\frac{2 \pi}{5}})$
gives the single-valued angular coordinates on $\C^5/\Z_5$; since the
$\theta_i$ are fifth roots of unity this fixes the coordinate
$\hat{\theta}_1=0$ and leaves the other $4$ coordinates as arbitrary
fifth roots of unity: I will refer to the $\hat{\theta}_i$ as the
``reduced'' angular coordinates on the orbifold.

We can represent the D-branes in $\C^5$ (graphically, as 5 copies of
$\C \simeq \R^2$, each of which contains a ray from the origin along
one of the fifth roots of unity) providing we remember the orbifold
condition that the target space is actually $\C^5/\Z_5$, in which one
of the rays is aligned with the positive real axis.

\begin{figure}[tbp]
\begin{center}
\epsfig{file=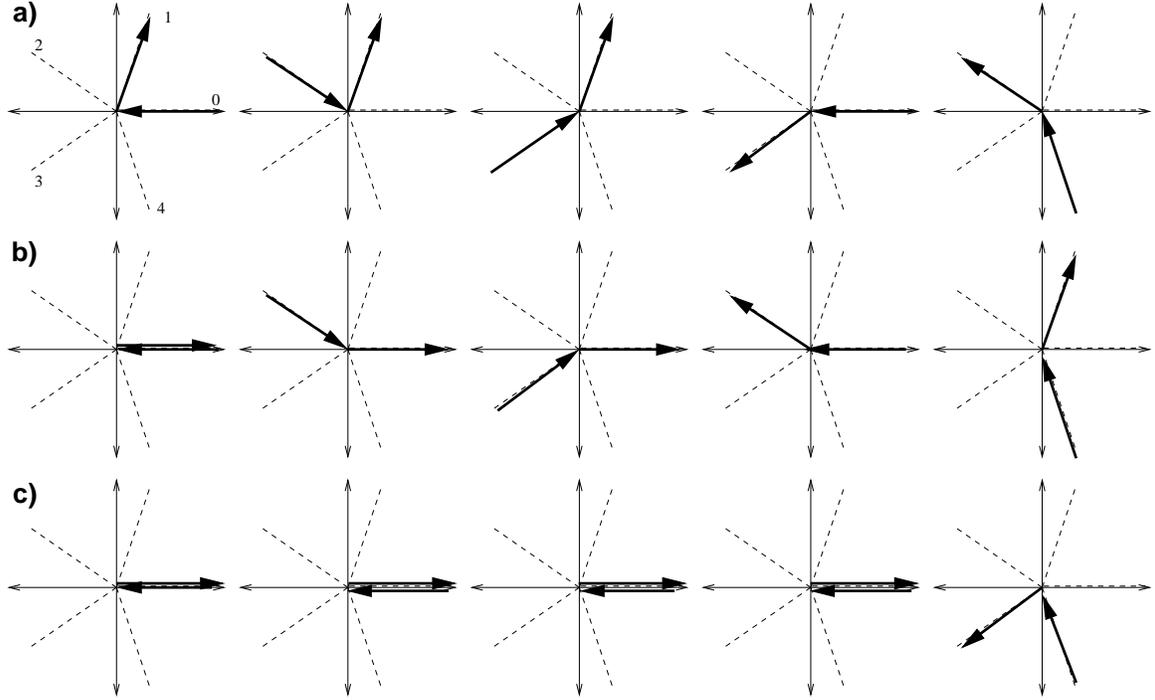}
\parbox{5.5in}{
\caption{(a) A typical pair of ``spokes'' in the five coordinate planes $\C$
  of $\C^5$: the incoming and outgoing rays in each coordinate plane
  are aligned with fifth roots of unity and parametrize the two
  half-branes in $\C^5$. (b) The image of the same spoke in
  $\C^5/\Z_5$: using the $\Z_5$ orbifold symmetry, which acts by a
  common phase rotation of each of the coordinates, we can
  ``collapse'' the pair in $z_1$, i.e.~align both the incoming and
  outgoing ray in the $z_1$ coordinate patch with the real axis, with
  a corresponding rotation of the other coordinates. (c) The image of
  the same spoke in the $\C^5/(\Z_5)^4$ mirror model; the symmetry
  action (\ref{mirrorz5}) can be used to align the first four pairs of
  rays with the real axis at the expense of a compensating rotation of
  the fifth coordinate.  The first four outgoing rays need to be
  rotated clockwise by a combined total of $1+1+1+3=6$ units to align
  them all; the orbifold symmetry causes the fifth to rotate
  counterclockwise by 6 units to the position shown.  Similarly, the
  incoming rays are rotated clockwise by $0+2+3+0=5$ units so the
  incoming fifth ray is rotated counterclockwise by 5 (and comes back
  to the same position).\label{spokes}} }
\end{center}
\end{figure}

In the mirror LG picture the orbifold group contains an additional
$(\Z_5)^3$: the combined $(\Z_5)^4$ symmetry can be taken to act in
the following way \cite{candelas}:
\begin{eqnarray}
\Z_5^{(1)} &:& (4, 0, 0, 0, 1) \nonumber \\
\Z_5^{(2)} &:& (0, 4, 0, 0, 1) \nonumber \\
\Z_5^{(3)} &:& (0, 0, 4, 0, 1) \nonumber \\
\Z_5^{(4)} &:& (0, 0, 0, 4, 1)
\end{eqnarray}
where the notation $(\alpha_1, \ldots, \alpha_n)$ means that the
symmetry acts on the coordinates by
\begin{equation}
(z_1, \ldots, z_n) \mapsto (\lambda^{\alpha_1} z_1, \ldots, \lambda^{\alpha_n} z_n), \lambda = e^{\frac{2 \pi \imath}{5}} .
\label{mirrorz5}
\end{equation}
We can use these symmetries to align four of the rays with the
positive real axis, with a compensating rotation of the fifth ray
leaving it along some other fifth root of unity.  See Figure
\ref{spokes} for an example of a pair of spokes in $\C^5$, and its
reduced image in the $\C^5/\Z_5$ orbifold and in the $\C^5/(\Z_5)^4$
orbifold of the mirror LG model.

As noted above, a single half-brane contains a boundary, and we
therefore need to study the possibilities for cancelling this
boundary to form a cycle.

Consider two half-branes specified by $(\hat{\theta}_1, \ldots,
\hat{\theta}_5)$ and $(\overline{\hat{\theta}}_1, \ldots,
\overline{\hat{\theta}}_5)$.  Along one of the boundary segments
$|z_i|^2=0$ of the half-branes, the $S^1_{(i)}$ fibre degenerates, and
so different values of $\hat{\theta_i}$ in fact represent the same
point.  If the other four $\hat{\theta_j}$ coordinates are distinct,
the two half-branes do not meet and therefore do not have a common
boundary segment there.

Since this is true over each boundary segment $|z_i|^2=0$, the only
way to produce two half-branes with common boundary is to take
$\hat{\theta}_i = \overline{\hat{\theta}}_i$ for all $i$.  These are
the {\it reduced} coordinates on the $\C^5/\Z_5$ orbifold; in terms of
the unreduced coordinates $\theta_i$ on $\C^5$ we can take a pair of
rays that differ by a common $\Z_5$ phase rotation.  The two sets of
rays will reduce to the same image in the orbifold, but since they are
only identified up to a $\Z_5$ rotation in $\C^5$ they are in a
twisted sector of the orbifold, and therefore the pair of half-branes
carries a $\Z_5$ topological charge which encodes the twist value of
the D-brane.  Heuristically speaking the pair of half-branes produce a
D-brane that comes in from infinity, twists around the orbifold fixed
point and heads off to infinity again along the same path in the
orbifold space (but a different path in the covering space).

For other values of $\hat{\theta}_i$, the two half-branes will only
meet at the origin, so they are topologically two copies of $(\R^+)^5$
touching at the origin inside $\C^5/\Z_5$.  The generic pair of
half-branes therefore still has a boundary.  We can complete this
submanifold to one without boundary by taking the ``doubled images''
of each half-brane under $\hat{\theta}_i \mapsto \hat{\theta}_i +
\pi$, together with a reversal of orientation to cancel the common
boundary segment \cite{holdiscs}.  This is the same thing as taking
$z_i \mapsto - z_i$, i.e.~adjoining another $(\R^+)^n$ along one
of the boundary hyperplanes; if we take $2^n$ such doublings we
complete $(\R^+)^n$ to $\R^n$ and obtain a submanifold without
boundary, which looks like two copies of $\R^n$ intersecting at the
origin inside $\C^n/\Z_n$.  Note that the doubling preserves the phase
of the holomorphic $5$-form (\ref{omega}), since both the shift in
$\hat{\theta}_i$ and the orientation reversal shift the phase by
$\pi$, leaving it invariant: the doubled half-brane is therefore still
\slag.

The fact that the generic pair of half-branes in the LG phase only
meet at the origin is not a problem for string theory, since we are
only interested in the infrared fixed point of the LSM.  Under RG flow
to the infrared the LSM coupling constant $g \rightarrow \infty$, so
the conformally-invariant string is confined to the classical vacuum
of the LSM (the orbifold fixed point).  The configuration space of the
infrared string is therefore the single point at the intersection of
the two half-branes, as expected for the LG models.

In the mirror model, because of the extra orbifold symmetry arising
from the Greene-Plesser construction there is no problem with boundary
cancellation of pairs of spokes, and in fact any two pairs of
half-branes will have a common boundary, without need to double the
geometry by taking $\theta_i \mapsto \theta_i + \pi$: using the
orbifold symmetry, we can collapse any four of the five pairs of rays
to align them with the real axis, and the fifth pair of rays will have
different values of $\hat{\theta}_i$.  Since different values of
$\hat{\theta}_i$ over any boundary $z_i=0$ still represent the same
point, and all of the other angular coordinates are equal by the
orbifold reduction process, the two boundaries are identified
automatically.

However, even when a pair of half-branes have a common boundary, we
still have to double the geometry in order to obtain submanifolds that
are \slag.  This will be apparent in two places below: in the
following paragraphs when we consider the construction of globally
\slag\ submanifolds from a pair of half-branes, and in the next
section when we consider the image of these submanifolds in the NLSM
phase, and their intersection with the quintic.

Given this construction of (special) Lagrangian submanifolds of the LG
orbifold target space, I will now derive their relationship to the
A-type rational boundary states of the Gepner model.

\begin{figure}[tbp]
\begin{center}
\epsfig{file=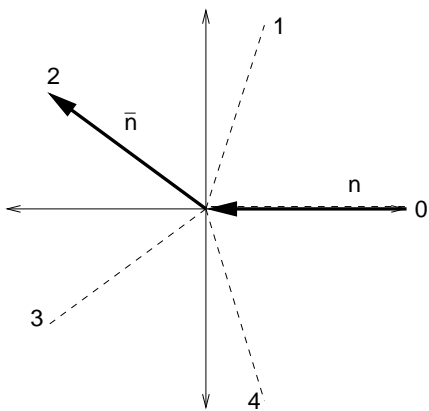}
\parbox{5.5in}{
\caption{The Lagrangian submanifolds corresponding to A-type rational boundary states are composed of a pair of rays aligned along roots of unity in each coordinate $z_i$ of the linear sigma model target space (fifth roots of unity for the quintic), with compatible orientation.  The submanifold represented here corresponds to the $L=1, M=2, S=1$ boundary state of the $(k=3)$ \ntwo\ minimal model; the A-type rational boundary states of the Gepner model are obtained by taking this class of boundary condition on each of the coordinates $z_i$.\label{spokefig}}
}
\end{center}
\end{figure}

Consider a pair of rays in a single coordinate $z_i$.  They are
labeled by $(k+2)^{\rm th}$ roots of unity where $n$ labels the
incoming ray and $\overline{n}$ labels the outgoing.  The
correspondence between these labels and the labeling $(L_i, M_i, S_i)$
of the A-type minimal model boundary states is as follows \cite{dmirror}:
\begin{eqnarray}
L_i &=& |\overline{n} - n| - 1 \nonumber \\
M_i &=& n + \overline{n} + \eta \nonumber \\
S_i &=& \hbox{sign(}\overline{n} - n\hbox{)} + \eta
\label{raylabels}
\end{eqnarray}
where $\eta=0,1$ for the R/R, NS/NS sectors respectively.  Since we
are interested in constructing BPS D-branes, which are objects with
R/R charge, I henceforth restrict to boundary states with $\eta=0$.
Geometrically, $L_i$ is related to the opening angle of the rays,
$M_i$ is related to the rigid rotation angle of the pair, and $S_i$
gives the orientation (see figure \ref{spokefig}).

This correspondence reproduces the field identification
(\ref{fieldid}) as follows.  There are two possible ways of
interpreting the opening angle of the pair of rays: we can either
start from $n$ and proceed counterclockwise around the circle to
$\overline{n}$ (giving $L_i = \overline{n} - n - 1$), or we can start
at $\overline{n}$ and proceed counterclockwise around the circle to
$n$, and then reverse the orientation to obtain the same submanifold.
Using the fact that the $n$ are defined mod $k+2$, this gives
\begin{eqnarray}
L'_i &=& (n + k + 2) - \overline{n} - 1 \nonumber \\
&=& k - (\overline{n} - n - 1) \nonumber \\
&=& k - L_i \nonumber \\
S'_i &=& S_i + 2 \nonumber \\
M'_i &=& (n+k+2) + \overline{n} \nonumber \\
&=& M + k + 2
\end{eqnarray}
as desired.

\begin{figure}[tbp]
\begin{center}
\epsfig{file=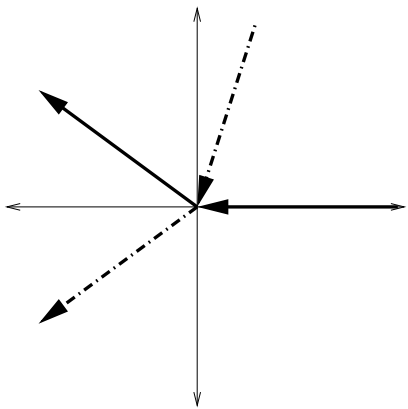}
\parbox{5.5in}{
\caption{Two intersecting pairs of spokes represented by the solid, and dashed pairs of rays.  Since the rays alternate as one proceeds anticlockwise around the circle they intersect positively.\label{interfig}}}
\end{center}
\end{figure}

Furthermore, the geometrical intersection of the pair of rays gives
the extended $SU(2)_k$ fusion coefficients \cite{dmirror}. Two
D-branes $(n, \overline n), (m, \overline m)$ which are each pairs of
rays will intersect with positive orientation if they ``overlap'',
i.e.~if the rays alternate or are paired between one D-brane and the
other as one proceeds counterclockwise around the circle (see figure
\ref{interfig}).
\begin{equation} n \leq m < \overline n \leq \overline m < n + k + 2
\label{interspokes}
\end{equation}
Some simple algebra \cite{dmirror} brings these inequalities into the
form (\ref{fusion}).  To obtain a negative intersection number we can
just reverse the orientation of one of the D-branes of a pair that
intersects positively; this can be expressed as a prefactor
$(-1)^{\frac{S-\overline S}{2}}$.  In other words, the intersection
form of two pairs of rays in a single coordinate is compactly
summarized by the extended $SU(2)_k$ fusion coefficients:
\begin{equation}
I_{\mbox{\small m.m.}}(L, \overline L, M, \overline M) = (-1)^{\frac{S-\overline S}{2}} N^{M-\overline M}_{L, \overline L}
\end{equation}

This can be extended to the intersection form of the A-type rational
boundary states of the Gepner model, which are built up from A-type
boundary states in each of the minimal model factors: the intersection
number of two sets of rays in $n$ coordinates is just the product of
the intersection number in each coordinate, and since the target space
is a $\Z_5$ orbifold we must sum over the intersections in the twisted
sectors, which all project to the same image in the orbifold space.
Since the $\Z_5$ acts by a common $M_j \mapsto M_j+2$ this is the same
as summing over all shifts of one of the sets of $M_j$ values by an
even integer, i.e.~relative $\Z_5$ rotations of the two pairs of
spokes.

Therefore,
\begin{eqnarray}
I_A &=& {\frac{1}{C}} (-1)^{\frac{{S-\overline S}}{2}} \sum_{\nu_0=0}^{K-1} \prod_{j=1}^r I_{\mbox{\small m.m.}}(L_j, \overline L_j, M_j + 2 \nu_0, \overline M_j ) \nonumber \\
&=& {\frac{1}{C}} (-1)^{\frac{S-\overline S}{2}} \sum_{\nu_0=0}^{K-1} \prod_{j=1}^r N^{2 \nu_0 + M - \overline M}_{L, \overline L}
\end{eqnarray}
where $C$ is an overall normalization constant, and we have identified
the $S_i$ values in the individual minimal models, as discussed in
section \ref{gepner}.  This is the desired intersection form
(\ref{intera}).  Upon dividing by the additional $(Z_5)^3$ orbifold
symmetry to get to the A-type intersection form on the mirror model,
one obtains the intersection form (\ref{interb}), which is also the
intersection form $I_B$ of the B-type states on the original manifold
(as expected by mirror symmetry).  Note that the presence of the
$SU(2)_k$ fusion rules in both A- and B-type intersection forms is
related to a geometrical description in terms of spokes.

Following \cite{quintic} the intersection forms $I_A, I_B$ can be
expressed as a polynomial in $\Z_5$ shift generators
\begin{equation}
g_i : M_i \mapsto M_i + 2
\end{equation}
The operator $g_i^{1/2}$ corresponds to a shift $M_i \mapsto M_i +
1$, which is needed when $L_j + \overline{L}_j = 1 \mbox{ mod } 2$ to
satisfy the relation (\ref{evengepner}).  The intersection form can be
expressed as a matrix where the entries are labelled by the $(M_j,
\overline M_j)$ values.

Algorithmically, the intersection matrix is built up by considering
the intersection number of two spoke pairs according to the rules
(\ref{interspokes}), summing over the $\Z_5$ shifts generated by $g_i$
(i.e.~the coefficient of each $g_i^a$ in the polynomial is given by
the intersection number of the first spoke pair and the second spoke
pair rotated by $g_i^a$.  For the A-type states the $g_i$ are subject
to the relation $\prod_{i=1}^5 g_i = 1$ which implements the
triviality of a common $\Z_5$ rotation, and for B-type states the
$g_i$ are all identified with a single $g$ by the $(\Z_5)^4$ symmetry.
Thus, the polynomial encodes the intersections of physically distinct
combinations of $M$ labels for a given set of $L_i$ labels.

For example, consider the intersection of two spokes with $L=1$ in the
$(k=3)$ minimal model, where the roots of unity of the two spoke pairs
are given by $(n, n+2 \mbox{ mod } 5)$ and $(m, m+2 \mbox{ mod } 5)$.
The intersections are summarized in Table \ref{mminter}:
\begin{table}[tbp]
\begin{tabular}{c|ll|cl}
Rel. shift&State 1&State 2&Intersection &\\ \hline
$1$ & $(0, 2)$ & $(0, 2)$ & $1$ & Since $0 \leq 0 < 2 \leq 2$ \\
$g$ & $(0, 2)$ & $(1, 3)$ & $1$ & $0 \leq 1 < 2 \leq 3 $\\
$g^2$ & $(0, 2)$ & $(2, 4)$ & $0$ & $0 \leq 2 \not< 2 \leq 4$ \\
$g^3$ & $(0, 2)$ & $(3, 0)$ & $-1$ & $0 \leq 0 < 2 \leq 3$ (Note orientation reversal) \\
$g^4$ & $(0, 2)$ & $(4, 1)$ & $-1$ & $0 \leq 1 < 2 \leq 4$
\end{tabular}
\begin{center}
\parbox{5.5in}{
  \caption{\small{The intersection number of two pairs of spokes, aligned along roots of unity given by $(n, n+2 \mbox{ mod } 5)$ and the same spoke pair rotated by $g^a$, i.e.~$(n+a, n+a+2 \mbox{ mod } 5)$, for the 5 values of $a$.  The intersection numbers are calculated according to the rules (\ref{interspokes}).}\label{mminter}}
}
\end{center}
\end{table}
Therefore
\begin{equation}
I_A(L=1) = (1 + g - g^3 - g^4)
\end{equation}
for the single minimal model.  Therefore for the A-type rational
boundary state of the full Gepner model the intersection matrix is:
\begin{equation}
I_A(L=\{11111\}) = \prod_{i=1}^5(1 + g_i - g_i^3 - g_i^4)
\end{equation}
subject to $\prod_{i=1}^5 g_i = 1$.  I will make use of this
intersection form in the next section.

The phase of the holomorphic $5$-form (\ref{omega}) (also known as the
``$U(1)$ grade'') of a half-brane $L$ is valued in the $\Z_5$ subgroup
of $U(1)$ and is given by the sum of the angles in each coordinate
plane.
\begin{eqnarray}
G = \hbox{Im log } \Omega^{(5)}|_L &=& \sum_{i=1}^5 \theta_i \nonumber \\
&=& {\frac{2 \pi}{5}} \sum_{i=1}^5 n_i
\label{spokegrade}
\end{eqnarray}
where $n_i$ label the roots of unity of the half-brane.  Note that it
is invariant under the $\Z_5$ orbifold symmetry.  When we take a
second half-brane $\overline{L}$ we must reverse its orientation in
order to have a compatible boundary orientation at the origin; this
shifts the assignment of the grade by $\pi$:
\begin{equation}
\overline{G} = \hbox{Im log } \Omega^{(5)}|_{\overline{L}} = \sum_{i=1}^5 (\overline{\theta_i} + \pi) = (\sum_{i=1}^5 \overline{\theta_i}) + \pi
\end{equation}
It is easy to verify that with this assignment of grades a D-brane
made up of two half-branes with opening angle $\pi$ in each coordinate
(i.e.~which is isomorphic to flat $\R^n$) will have a constant grade
everywhere; and conversely a half-brane and the same half-brane taken
with opposite orientation (i.e.~the antibrane to the first) have a
relative grade of $\pi$.

This assignment of grades forces us to take at least one of the
coordinates of the half-brane to its doubled image in order to obtain
a pair of half-branes with the same grade: if all 5 $\theta_i$ angles
are valued in fifth roots of unity for both incoming and outgoing
half-branes, then there will be an extra shift of $5 \pi \simeq \pi$
between the two grades, and for models with $k=\mbox{odd}$ there is no
way to make the two grades equal (since $\pi$ is not a $(k+2)^{\rm th}$ root
of unity).  This is remedied by taking an odd number of the $\theta_i$
to their doubled image $\theta_i \mapsto \theta_i + \pi$, which
causes the two grades to be valued in the same $\Z_5$ subgroup of
$U(1)$ so they can potentially be equal.  Thus, in each minimal model
($z_i$ coordinate) the boundary conditions are isomorphic to those
constructed in \cite{dmirror}, but there are additional constraints on
how the boundary conditions on each of the $z_i$ can be glued together
to form supersymmetric boundary conditions for the full Gepner model
($\C^5/\Z_5$ orbifold).

The grade of the submanifold is the same in both phases of the LSM,
because the $\Z_5$ action on the roots of unity becomes part of the
projective action on the coordinates of $\P^4$.  Therefore D-branes
that are mutually \slag\ in one phase will still be in the other.
This is to be expected because the \slag\ submanifolds do not decay
under variation of \Kah\ structure \cite{stability}.  Their stability
depends only on the complex structure of the \CYM\ and not the \Kah\ 
structure (except through the Lagrangian condition), and we have fixed
the 101 complex structure moduli of the quintic to 0 throughout.

The condition for two half-branes $\{\theta_i\}$, $\{\overline{\theta}_i\}$ to
preserve the same A-type \SUSY\ is therefore:
\begin{eqnarray}
\sum_{i=1}^r \overline{\theta}_i &=& \sum_{i=1}^r \theta_i \nonumber \\
\Leftrightarrow \sum_{i=1}^r (\overline{n}_i - n_i) &=& 0 \nonumber \\
\Leftrightarrow \sum_{i=1}^r L_i &=& r \nonumber \\
 & \equiv & 0
\label{lsusy}
\end{eqnarray}
using the labeling identification $L_i = \overline{n}_i - n_i - 1$,
and where the last equivalence is true for the quintic since $k+2 = r
= 5 \equiv 0 \mbox{ mod } 5$, but may not be true in a more general
model.

D-branes with $\sum_i L_i \neq r$ are not \slag\ since the two
half-branes have different grades, and we expect them to flow to
another state in the same homology class that is \slag.  For the
quintic (and the $(k=1)^3$ torus model) every pair of spokes has a
\slag\ in the same homology class that is also another pair of spokes
(see Appendix \ref{slagproof} for the proof of this result).  For more
general Gepner models, construction of the \slag\ appears to be more
subtle, because some of the spoke pairs may fail to have a \slag\ 
representative that is also a spoke pair.  Since the spoke pairs
reproduce the topological properties of the A-type boundary states,
they are still good topological representatives of this unknown \slag,
but I do not presently know how to explicitly construct it for those
that fail.

Comparison of (\ref{raylabels}), (\ref{spokegrade}) and (\ref{lsusy})
shows that for \slag\ spokes of the quintic, the grade
(\ref{gepnergrade}) of the Gepner model boundary states from CFT
reduces to

\begin{equation}
\sum_{i=1}^r \frac{M_i}{k_i+2} = \sum_{i=1}^5 \frac{n_i + \overline{n_i}}{5} = \frac{2}{5} \sum_{i=1}^5 n_i = \frac{G}{\pi}
\end{equation}
i.e.~it is in agreement with the geometrical grade of the \slag\ (the
phase of the holomorphic 5-form).

\subsection{D-branes in the large volume limit}
\label{geombr}

Having characterized the properties of the spokes in the LG phase,
where they are related to the rational boundary states of the Gepner
model, I now turn to their properties in the geometrical phase $r>0$.

Recall that the transition from the LG phase to the NLSM phase of the
quintic blows up the orbifold fixed point at the origin of $\C^5/Z_5$
into a $\P^4$, which becomes the zero section of the line bundle
$\linefour$.  The vacuum submanifold of the linear sigma model in this
phase is the the quintic hypersurface in $\P^4$, and we will see that
the 5-dimensional \slag\ submanifolds of $\linefour$ (pairs of spokes)
intersect this hypersurface to form 3-dimensional \slag\ submanifolds
of the quintic.

Also recall from section \ref{toric} that the Ricci-flat metric on
$\linen$ is the Calabi metric, which has a copy of the Fubini-Study
metric (\ref{fubinistudy}) on the $\P^{n-1}$ at $w=0$:
\begin{equation}
ds^2 = {\frac{1}{\rho}} \left( dy_i dy^i - {\frac{1}{\rho}} y_i dy^i y_{\bar j} dy^{\bar j} \right)
\end{equation}
where $\rho = 1 + y_i y^i$, and $y_i$ are inhomogeneous coordinates on
the $\P^{n-1}$ in a coordinate patch ${\cal U}_n$.  $\P^{n-1}$ may be
described topologically by a copy of $\C^{n-1}$, plus a $\P^{n-2}$ at
infinity that compactifies the space.  In a local coordinate patch we
do not see the $\P^{n-2}$ at infinity and the patch is diffeomorphic
to $\C^{n-1}$.

We need to find the change of coordinates from the coordinates induced
from $\C^{n+1}$ on the line bundle, to the inhomogeneous coordinates on
$\P^{n-1}$.  Consider a vertex of the $(n-1)$-simplex that is the
toric base of the $\P^{n-1}$ in the induced metric from $\C^{n+1}$.
There are $n-1$ lines meeting at the vertex, and the face opposite to
the vertex is an $(n-2)$-simplex.  The Fubini-Study metric on
$\P^{n-1}$ effectively stretches out the $n-1$ lines meeting the
vertex to infinite coordinate distance; the opposite $(n-2)$-simplex
is pushed off to infinity.  The $(n-2)$-simplex is the base of a
$\P^{n-2}$ when we include the (degenerate) $T^{n-1}$ fibres, and if
we delete it we are left with a non-compact space which is
diffeomorphic to $\C^{n-1}$ since each of the coordinates along the
lines meeting our vertex also comes with an $S^1$ fibre (recall the
toric construction of $\C^n$ in section \ref{toric}).  We can repeat
this construction for any of the $n$ vertices of the $(n-1)$-simplex;
these are the $n$ coordinate charts of the $\P^{n-1}$.

We can parametrize the simplex at $|z_{n+1}|^2 = 0$ by the $n-1$
coordinates $(|z_1|^2, \ldots, |z_{n-1}|^2)$.  The remaining
coordinate $|z_n|^2$ depends on the first $n-1$ according to
\begin{equation}
|z_n|^2 = r - \sum_{i=1}^{n-1} |z_i|^2
\end{equation}
in order to satisfy the D-term constraint (\ref{linedterm}).  We can
obtain the other coordinate patches by choosing a different set of
$n-1$ independent coordinates to parametrize the simplex.  Therefore,
the $\C^n$ coordinates $z_i$ at $|z_{n+1}|^2 = 0$ are similar to the
usual homogeneous coordinates on $\P^{n-1}$, except that the $|z_i|^2$
range from $0$ to $r$.

The change of variables into the Fubini-Study coordinates is given by:

\begin{eqnarray}
y_i &=& \frac{z_i}{z_n} \nonumber \\
\Leftrightarrow |y_i|^2 &=& \frac{|z_i|^2}{{r - \sum_{i=1}^{n-1} |z_i|^2}} \\
y_i &=& |y_i|^2 e^{\imath (\theta_i-\theta_n)}
\end{eqnarray}
which are just the usual projective coordinates on $\P^{n-1}$.  Note
that since the angles are shifted by roots of unity the projected
spokes again look like spokes in the coordinate patch, except that
they are only spokes in $n-1$ coordinates; the remaining coordinate is
pushed off to infinity in the coordinate patch on $\P^{n-1}$, but it
is visible by considering the image of the submanifold in two
different coordinate patches.

In projective coordinates the quintic polynomial becomes
\begin{equation}
1 + \sum_{i=1}^4 (y_i)^5 = 0
\label{projquintic}
\end{equation}
If the submanifolds are parametrized along $n^{\rm th}$ roots of
unity, then this equation has no solution on the submanifold, because
$y_i^5 \in \R^+$.  Therefore, in order for the D-branes to intersect
the quintic hypersurface in $\P^{n-1}$ we must take one or more of the
$\theta_i \mapsto \theta_i + \pi$ to introduce a relative minus
sign into one of the terms in (\ref{projquintic}).  This is the same
prescription that was required to cancel the boundary of two piecewise
\slag\ submanifolds and to obtain globally \slag\ submanifolds from
the pair.

This construction reproduces an old construction of 3-cycles on the
mirror quintic from \cite{candelas}, which are used to compute the
periods of the holomorphic 3-form.  The construction was further
analyzed in \cite{periods}\footnote{There is a slight
  over-generalization in the discussion of these 3-cycles in the
  appendix to that paper, which implies that an arbitrary pair of
  spokes on a general \CYM\ will have a common boundary.  As shown in
  section \ref{lg} this is in fact only true for the Greene-Plesser
  mirror where there is additional orbifold symmetry.  This correction
  does not affect the results of that paper, since they only use the
  construction to study periods of the mirror manifold.}, where the
3-cycles were termed ``spokes'', and in \cite{semiperiods}.  Those
papers were concerned with the intrinsic geometry of the compact \CYM\ 
itself, however the analysis naturally fits into the LSM framework
discussed in this paper since the holomorphic 3-form on the compact
\CYM\ is obtained by pullback from $\C^{n+r}$ via the construction
outlined in section \ref{toric}, so A-type D-branes of the LSM descend
to A-type D-branes of the compact \CYM\ in the infrared.  The results
of \cite{candelas} were also used in \cite{quintic} to study the
B-type rational boundary states of the quintic at the Gepner point and
at large volume, which are mirror to A-type states of the mirror
quintic.  Those results will be analyzed in more detail in the
following section.

Topological properties of the spokes such as intersection numbers
should not change on transition from the LG phase to the NLSM phase,
since from the point of view of the LSM we are considering
submanifolds defined by constant $\theta_i$, and the transition
between the two phases is implemented by translating the defining
hyperplane (\ref{hyperplane}) away from the origin in $\C^6$ to
truncate the ``tip'' of the toric base space into a simplex (see
section \ref{toric}). In other words the transition does not modify
the fibre coordinates and the passage from LG to NLSM phase simply
``chops off the tip'' of the 5-dimensional half-branes.

This accords with the observation in \cite{quintic} that the
intersection form of the $L=(11111)$ A-type states in the $(k=3)^5$
Gepner model is the same as the intersection form of certain \slag\ 
3-cycles on the quintic (the 625 $\R\P^3$s constructed in section
\ref{cym}), however there is an important difference which I will now
discuss.

Recall from section \ref{cym} that the $\R\P^3$ \slag s were
constructed as the fixed-point set of a real involution, and their
image in each of the coordinates $z_i$ of $\C^5$ is a straight line
aligned along the $\omega_i^{\rm th}$ fifth root of unity (i.e.~two
rays through the origin with opening angle of $\pi$).  However, the
construction of the $L=(11111)$ spokes gives a submanifold of the
quintic that is ``bent'' and has an opening angle of $\frac{4 \pi}{5}$
in each of the $z_i$ planes.  It is also a \slag\ submanifold since
the two half-branes have the same grade, as discussed above.

As discussed in section \ref{lg} the spoke construction correctly
reproduces the intersection form $I_A$ of the A-type rational boundary
states and is equal to
\begin{equation}
I_A(11111)=\prod_{i=1}^5(1 + g_i - g_i^3 - g_i^4)
\end{equation}
The intersection form of the $\R\P^3$s was calculated in \cite{quintic} to be
\begin{equation}
I_{\R\P^3}=\prod_{i=1}^5(g_i + g_i^2 -g_i^3 - g_i^4)
\end{equation}
which is equal to $I_A$ up to the relation $\prod_{i=1}^5 g_i = 1$ and
an overall minus sign (which presumably comes from a relative change
of orientation between the conventions used to define the $\R\P^3$s
and the $L=(11111)$ spokes).  Therefore, we have two distinct \slag s
with the same intersection numbers.

In terms of the gauge theory living on the world-volume of the
D-brane, stability of the D-brane (i.e.~existence of a stable vacuum)
is governed by the D-terms of the gauge theory, whereas the moduli
space of deformations of the D-brane is determined by the F-flatness
conditions on the world-volume superpotential:
\beq
\label{fflat}
\frac{\partial W}{\partial \psi} = 0
\eeq
where $\psi$ are the massless chiral fields of the world-volume
theory.  In other words, flat directions of the superpotential
correspond to exactly marginal deformations.

For A-type D-branes, the world-volume D-terms depend only on the \cs\ 
moduli of the \CY, and similarly the superpotential (and therefore the
moduli space of deformations of a given stable D-brane) depends only
on the \Kah\ moduli (as usual, the role of \Kah\ and \cs\ moduli are
interchanged by mirror symmetry, i.e.~for the mirror B-model).  Away
from the large volume limit the world-volume superpotential receives
corrections from open string instantons, since the area of the open
string world-sheet is measured by the \Kah\ form.

McLean \cite{mclean} showed that the moduli space of deformations of a
compact \slag\ submanifold $L$ is (locally) a smooth manifold of
dimension $b_1(L)$ (the first Betti number of $L$).  In particular,
\slag\ submanifolds with vanishing first Betti number are rigid.  This
is the case for the \slag\ submanifolds of the quintic under
discussion, since they are diffeomorphic to $\R\P^3$ which has
$H_1(\R\P^3, \Z) \simeq \Z_2$.  Since we have two rigid \slag s in the
same homology class, the 0-dimensional moduli space has (at least) two
components.  In fact, all of the \slag\ submanifolds associated to
rational boundary states of the quintic are topologically $\R\P^3$, by
a piecewise version of the argument given in section \ref{cym}.
Therefore {\it all} the A-type D-branes associated to A-type rational
boundary states of the quintic are rigid in the large volume limit.

The instanton corrections to the superpotential may be thought of as
``stringy'' modifications to the classical deformation theory of the
\slag\ submanifolds.  In other words, McLean's theorem is only true
perturbatively in $\alpha'$ and may be violated non-perturbatively in
$\alpha'$ by instanton corrections \cite{openinst,
  mirroropen}\footnote{I am grateful to I.~Brunner for a discussion on
  this point.}.

In the large volume limit, marginal operators of the world-volume
gauge theory correspond to geometrical deformations of the \slag\ 
submanifold and the flat $U(1)$ gauge bundle living on it, which are
both classified by $H_1(L, \Z)$ \cite{syz} and pair to form complex
moduli fields \cite{mirroropen} (when this group is finite there are
$|H_1(L, \Z)|$ distinct choices of flat $U(1)$ bundle \cite{joyce}).
Away from the large volume limit, open string instanton contributions
to the superpotential are enumerated by a choice of 1-cycle of the
\slag\ upon which the open string world-sheet ends (as well as a
2-cycle of the bulk \CY\ around which the interior of the world-sheet
wraps) \cite{holdiscs}.  In particular, for the rigid \slag s under
discussion, $H_1(L, \Z) \simeq \Z_2$ and the instanton contribution
dramatically simplifies (an open string worldsheet can only wind once
around the single 1-cycle to give a nontrivial contribution).

Since there are no deformations of these \slag s in the large volume
limit (i.e.~massless fields), the only way they can arise at the
Gepner point is due to instanton effects.  The counting of massless
boundary fields at the Gepner point was described in \cite{quintic}.
One finds that the only A-type rational boundary states of the quintic
which posess massless fields are the $L=(11111)$ states (which have
one), and all other boundary states have no massless fields in their
spectrum.

This massless field may have an instanton-generated superpotential.  A
superpotential was postulated in \cite{quintic} of the form
\begin{equation}
W = \psi^3 + \psi \phi
\end{equation}
where $\psi$ is the massless boundary field and $\phi$ is the (bulk)
\Kah\ modulus.  The cubic term in this superpotential was calculated
explicitly in \cite{gepnersuper} and was indeed found to be
non-vanishing.  The term linear in $\psi$ was not calculated
explicitly in that paper, but it is not forbidden by selection rules
and is therefore likely to also be non-vanishing.  Thus away from the
Gepner point ($\phi = 0$) the superpotential has two distinct vacua
corresponding to the distinct gauge bundles on a \slag\ in the
$L=(11111)$ homology class, and these vacua combine and become
degenerate at the Gepner point.  Therefore all of the A-type rational
boundary states of the quintic are rigid at the Gepner point as well
as in the large volume limit.

The story for the A-type states of the mirror quintic is more
complicated, since these \slag s may have $b_1 > 0$ due to
identifications under the Greene-Plesser orbifold action.  Indeed, the
boundary spectrum of these D-branes at the Gepner point generally
contains large numbers of massless fields \cite{quintic}, and explicit
computations in the B-model \cite{quinticW} shows that these fields
often remain exactly marginal at the Gepner point (i.e.~the flat
classical superpotential $W = 0$ is not completely lifted by the
instanton contributions).

\subsection{Relationships within the charge lattice}
\label{chargelatt}

I now turn to the question of relationships between the rational
boundary states, from the point of view of the D-branes of the linear
sigma model constructed in section \ref{lg}.  I will show that this
construction correctly reproduces the numerous relationships which
exist between the large-volume homology classes of the rational
boundary states; by exploiting the ``spoke'' structure of the D-branes
this reduces to a simple problem involving the addition of
polynomials.  This serves as a nontrivial check of the construction,
and the discussion may be a useful starting point for describing the
dynamics of tachyon condensation of an unstable pair of boundary
states into a bound state.

The A- and B- type rational boundary states of the Gepner model
associated to the quintic were studied in \cite{quintic} and were
related to geometrical objects in the NLSM phase using results from
\cite{candelas} (the main results from \cite{quintic} on A-type states
were reproduced in the previous two sections).  By essentially
ignoring the problem of stability of the B-type states as they are
transported through \Kah\ moduli space, a set of large-volume
topological invariants (Chern classes, or equivalently, the D-brane
Ramond-Ramond charges) can be associated to these B-type D-branes
using an algorithmic procedure.

The analysis of \cite{quintic} mostly focused on B-type states since
these are both fewer in number for the quintic and easy to associate
to known geometrical objects (such as bundles and sheaves on $\P^4$);
however since the B-type rational boundary states map under mirror
symmetry to the A-type rational boundary states of the mirror model we
can hope to reproduce these results using the LSM construction of
A-type D-branes of the mirror model.

Computing the spectrum of stable B-type D-branes at a generic point in
\Kah\ moduli space is a difficult task because the B-type D-branes can
decay under \Kah\ deformations away from the deep interior points
where the spectrum is known \cite{stability}.  The instability of
B-type D-branes under variation of \Kah\ structure is due to instanton
effects and is difficult to study directly (it is formulated in terms
of ``$\Pi$-stability'' of vector bundles \cite{stability}, and the
formalism of the derived category of coherent sheaves \cite{dcat,
  decat0, dstabmon}); however under mirror symmetry this is mapped
into the geometrical problem of stability of \slag\ submanifolds under
variation of complex structure.  It is known \cite{joyce} that there
are ``walls'' (hypersurfaces of real codimension one) in the \cs\ 
moduli space of \CY\ 3-folds; as one deforms towards such a wall, a
family of \slag\ submanifolds becomes singular and it can cease to
exist on the other side of the wall.

Thus, we also expect there to be stability problems in transporting
A-type D-branes from the mirror to the Gepner point $\psi=0$ to the
large complex structure limit $\psi \rightarrow \infty$, but we can
study the topology of such objects in the same sense as in
\cite{quintic}.

The spoke pairs constructed in the previous section descend to \slag\ 
submanifolds of the compact \CY\ when the complex structure moduli are
fixed to zero: away from this point the singular, V-shaped spokes are
deformed and become smoothed out.  This process was studied briefly in
\cite{candelas, dmirror}, and it may be possible to obtain a more
complete understanding of the stability of BPS D-branes by studying
complex structure deformations of A-type D-branes in more detail.

The spectrum of D-brane charges of the B-type Reckangel Schomerus
states was computed in \cite{quintic}.  They form an overcomplete
basis for the D-brane charge lattice of the quintic, but they are not
an integral basis because the D0 charge of the B-type states only
occurs in multiples of 5.  This is a generic feature of the B-type
rational boundary states and is true even for the simplest Gepner
models, which are associated to 2-tori.

The complete spectrum of A-type D-branes on a torus $T^2$ is given by
circles $S^1$ with all possible integer winding numbers; however
(roughly speaking) the rational boundary states only correspond to the
D-branes with unit winding numbers, and not the higher winding cycles.
Under mirror symmetry (T-duality) they become even-dimensional cycles
with D0 charge given by multiples of some integer (multiples of 3 for
the $(k=1)^3$ torus model).  The D0 charge comes from the complex
structure parameter $\tau$ of the torus: a D1-brane wrapping along the
lattice vector $\tau$ dualizes into a D-brane system with flux coming
from the angle \cite{dlectures}.

In other words, the A-type rational boundary states are an integral
basis for the middle-dimensional homology, but under mirror symmetry
the B-type states at the Gepner point are not an integral basis for
$H_0$ but form a sublattice of finite index within it.  This can be
explained by the additional orbifold symmetry that exists in the LG
phase: for the quintic we cannot consider just a single D0-brane in
the LG orbifold model, but must consider its images under the $\Z_5$
symmetry as well.  In the NLSM phase there is no orbifold symmetry,
and a single D0-brane can potentially be BPS (as expected, since in
the limit of large volume the \CY\ becomes approximately flat, and a
single D0-brane is BPS in flat space).

The main result of this paper (see also \cite{stringycy}) is that the
rational boundary states descend from certain submanifolds of the
linear sigma model in both phases.  Therefore it is no great surprise
that they produce objects with multiple D0 charge even in the NLSM
phase: for example a generic curve in $\P^4$ (i.e.~a D-brane inherited
from the non-compact linear sigma model target space) will intersect
the quintic in 5 points, giving a D-brane with D0 charge of 5.

The rational boundary states are only {\it generators} for the
homology lattice, and it seems hard in general to construct a larger
class of boundary states purely within the Gepner models\footnote{See
  \cite{wound} for a construction of the higher-winding boundary
  states in the $T^2$ models; unfortunately this construction does not
  seem to generalize to the higher-dimensional Gepner models.}, but
since we have associated these boundary states to a certain subclass
of the D-branes in the LSM it is clear that a more general
construction should be possible: presumably these more general
D-branes will give rise to states of higher D-brane charge.  I will
discuss the possibilities for constructing more general Gepner model
boundary states in section \ref{newbranes}.

I present the complete list of D-brane (Ramond-Ramond) charges of the
B-type rational boundary states on the quintic in Table
\ref{chargetable}.  These were obtained using the algorithm in
\cite{quintic}\footnote{The partial list of D-brane topological
  invariants in that paper is given in terms of the rank and Chern
  numbers $\hbox{Ch}_n$ of the vector bundles instead of the D-brane
  charges.}.  In order to look for relationships between the A-type
D-branes (and therefore the B-type D-branes on the mirror) it is
convenient to encode the submanifolds as certain polynomials.

Recall that for the models under consideration the D-branes in the LSM
description are classified by a pair of rays aligned along $n^{\rm
  th}$ roots of unity (with compatible orientation) in each coordinate
$z_i$, $i = 0, \ldots, n-1$, where not all of the choices in each of
the $z_i$ coordinates are distinct due to the orbifold symmetry of the
target space which also acts by some combination of $\Z_n$ rotations.

\begin{table}[tbp]
\begin{center}
\begin{tabular}{r|r|rrrrlr|r|rrrr}
$\{L_i\}$ & M & Q6 & Q4 & Q2 & Q0 &\hspace{1in}       &$\{L_i\}$ & M & Q6 & Q4 & Q2 & Q0 \\ \cline{1-6} \cline{8-13}
00000 & 5 & 1  & 0  &  0 & 0      &                   &    00111 & 8 & -1 & 0  & 5  & 5  \\
      & 7 & -4 & -1 & -8 & 5      &                   &          & 0 & 11 & 6  & 38 & -20\\
      & 9 & 2  & 0  & 5  & 0      &                   &          & 2 & 6  & 3  & 14 & -15\\
      & 1 & 4  & 3  & 14 & -10    &                   &          & 4 & -6 & -4 & -27& 10 \\
      & 3 & -3 & -2 & -11& 5      &                   &          & 6&-10& -5 & -30& 20 \\ \cline{1-6} \cline{8-13}

00001 & 6 & -3 & -1 & -8 & 5      &                   &    01111 & 9 & 10 & 6  & 43 & -15\\
      & 8 & -2 & -1 & -3 & 5      &                   &          & 1 & 17 & 9  & 52 & -35\\
      & 0 & 6  & 3  & 19 & -10    &                   &          & 3 & 0  & -1 & -13& -5 \\
      & 2 & 1  & 1  & 3  & -5     &                   &          & 5 & -16& -9 & -57& 30 \\
      & 4 & -2 & -2 & -11& 5      &                   &          & 7 & -11& -5 & -25& 25 \\ \cline{1-6} \cline{8-13}

00011 & 7 & -5 & -2 & -11& 10     &                   &    11111 & 0 & 27 & 15 & 95 & -50\\
      & 9 & 4  & 2  & 16 & -5     &                   &          & 2 & 17 & 8  & 39 & -40\\ 
      & 1 & 7  & 4  & 22 & -15    &                   &          & 4 & -16& -10& -70& 25 \\ 
      & 3 & -1 & -1 & -8 & 0      &                   &          & 6 & -27& -14& -82& 55 \\ 
      & 5 & -5 & -3 & -19& 10     &                   &          & 8 & -1 & 1  & 18 & 10 \\
\end{tabular}
\parbox{5.5in}{
\caption{\small{Large-volume D-brane (Ramond-Ramond) charges of the B-type rational boundary states,
    computed using the procedure in \cite{quintic}.  Permutations of
    the set of $L_i$ values have the same charges, and performing a
    ``field identification'' on $n$ of the $L_i$ to interchange $L_i=0
    \leftrightarrow L_i=3$ or $L_i=1 \leftrightarrow L_i=2$ (which
    also shifts the $M$ labels) introduces an overall sign of
    $(-1)^n$.  Therefore the remaining combinations of $\{L_i\}$ not
    listed here also have these charges up to an overall minus sign.}
\label{chargetable}}
}
\end{center}
\end{table}

For each independent coordinate, we introduce a $\Z_n$ rotation
generator $R_i$ (i.e.~satisfying $R_i^n = 1$) which acts on the
positive real axis in $\C$ to produce an outgoing ray aligned in the
$e^{2 \pi \imath/n}$ direction (recall that the \slag\ submanifolds we
are considering look like copies of $(\R^+)^n$, i.e.~rays $\R^+$ in
each coordinate.  Reversing the orientation of the ray (to obtain the
incoming ray) is represented by taking $- R_i$ instead of $R_i$.

In this formalism, the A-type D-branes constructed in section \ref{lg}
are represented by completely factorizable polynomials which can be
written as the product of $n$ terms:

\begin{equation}
P(R_1, \ldots, R_n) = \prod_{i=1}^n (R_i^{k_i} - R_i^{{k_i}'})
\end{equation}
which corresponds to rays along the $(k_i, {k_i}')$ roots of unity in
the $i^{\rm th}$ coordinate.  The labelling identification
(\ref{raylabels}) can be used to translate back and forth between the
polynomials, the $(L_i, M_i, S_i)$ labels of the boundary states, and
the geometrical image of the D-brane.  Note that these polynomials are
different from the polynomials discussed in section \ref{lg}; those
polynomials encoded the intersection of two boundary states; the
polynomials currently under discussion describe the geometrical image
of a single D-brane.

The $\Z_5$ symmetries of the quintic and its mirror are manifested as
relations between the generators $R_i$: for the quintic the generators
satisfy
\begin{equation}
\prod_{i=1}^5 R_i = 1
\label{rotrel1}
\end{equation}
while for the mirror the $(\Z_5)^4$ symmetry (\ref{mirrorz5}) implies
\begin{eqnarray}
R_i^4 R_5 &=& 1,\ i=1, \ldots 4 \nonumber \\
\Leftrightarrow R_i^5 R_5 &=& R_i \nonumber \\
\Leftrightarrow R_5 &=& R_i
\label{rotrel2}
\end{eqnarray}
i.e.~the generators are all identified and we are left with a
polynomial in a single generator.

In this language, it is clear how one can look for relationships
between the D-branes corresponding to rational boundary states: find
two factorizable polynomials (rational boundary states) that, when
added together, produce a third polynomial that is also factorizable
(another rational boundary state), up to the relations (\ref{rotrel1})
or (\ref{rotrel2}).

For example, consider the two A-type D-branes on the mirror quintic
represented by
\begin{eqnarray}
P_1 &=& (R - 1)^5  \nonumber \\
P_2 &=& (R - 1)^4 (R^2 - R)
\end{eqnarray}
which are equivalent by mirror symmetry to B-type rational boundary
states on the quintic, in this case two states in the $L=(00000)$
orbit with $M=(1 \cdot 5) + (0 \cdot 5)=5$ and $M=(1 \cdot 4 + 2) + (0
\cdot 4 + 1) = 7$ respectively.  Under addition:
\begin{eqnarray}
P_{(1+2)} = P_1 + P_2 &=& (R-1)^4 \left[ (R - 1) + (R^2 - R) \right] \nonumber \\
&=& (R-1)^4 (R^2 - 1)
\label{poly1}
\end{eqnarray}
which is the polynomial corresponding to the state $\{L=(00001), M=6\}$.
Referring to Table \ref{chargetable} it is seen that addition of the
D-brane charges is indeed satisfied:
\begin{equation}
\begin{array}{ccccc}
\{ L=(00000), M=5 \} &+& \{ L=(00000), M=7 \} &=& \{L=(00001), M=6\} \nonumber \\
(1, 0, 0, 0) &+& (-4, -1, -8, 5) &=& (-3, -1, -8, 5)
\end{array}
\end{equation}

A more complicated example is:
\begin{eqnarray}
P_1 &=& (R-1)^3(R^2-1)^2 \nonumber \\
P_2 &=& (R-1)^2(R^2-1)^2(R^4-R^2) \nonumber \\
P_{(1+2)} &=& (R-1)^2(R^2-1)^2 \left[ (R-1) + (R^4-R^2) \right] \nonumber \\
 &=& (R-1)^2 \left( (R-1)^2 (R+1)^2 \right) \left[ (R-1) + (R^2-1)R^2 \right] \nonumber \\
 &=& (R-1)^4 \left[ (R+1)^2(R-1) + (R+1)^3(R-1)R^2 \right] \nonumber \\
 &=& (R-1)^4 (R^5 - R^3)
\end{eqnarray}
which corresponds to the charge relation
\begin{equation}
\begin{array}{ccccc}
\{L=(00011),M=7\} &+& \{L=(00111), M=2\} &=& \{L=(00001), M=2\} \nonumber \\
(-5,-2,-11,10) &+& (6, 3, 14, -15) &=& (1,1,3,-5)
\end{array}
\end{equation}
The other relationships within the charge lattice may be obtained
similarly.

\begin{figure}[tb]
\begin{center}
\epsfig{file=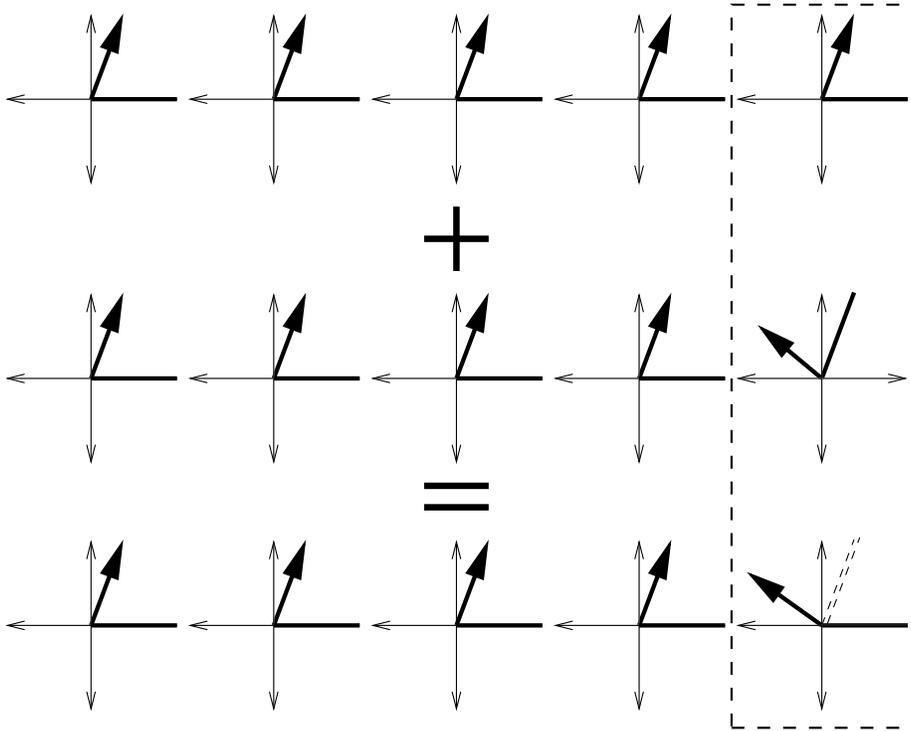}
\parbox{5.5in}{
\caption{Tachyon condensation of the $L=\{00000\}, M=5$ state with the $L=\{00000\}, M=7$ state to give the $L=\{00001\}, M=6$ state, as described in (\ref{poly1}) in terms of polynomials.  The spoke pairs in the first 4 coordinates of $\C^5$ are the same, but in the fifth coordinate the outgoing ray of the first state is anti-aligned with the incoming ray of the second state.  Addition of the two states is achieved by superposition of the pairs of spokes, and the anti-aligned rays are erased (shown as dashed lines in the above figure) to give the endpoint of the condensation process.\label{erasurefig}}}
\end{center}
\end{figure}

Translating back to the geometrical language, this again has a simple
interpretation; two rational D-branes that combine in such a way to
produce another rational D-brane contain a pair of rays which align
with opposite orientation, with the other rays in those coordinates
being distinct, and in the other coordinate planes both pairs of rays
coincide with the same orientation (see Figure \ref{erasurefig}).  The
pair of rays with opposite orientation define a homologically trivial
submanifold and can be collapsed to give a representative element of
the same homology class (which is the sum of the homology classes of
the original two D-branes).  This is the D-brane corresponding to the
sum of the two polynomials.

In physical language, the process of erasing anti-aligned D-brane
segments is very reminiscent of tachyon condensation of a coincident
D-brane and anti-D-brane pair \cite{sen1, sen2, relevance,
  braneantibrane}.  We consider two D-branes that are individually BPS
and therefore stable (i.e.~each of them preserve a particular linear
combination of the \ntwo\ \SUSY\ generators of the target space), but
taken together they do not preserve supersymmetry (i.e.~they do not
preserve the {\it same} linear combination of \SUSY\ generators, so
there is no combination of supersymmetries under which the combined
system is invariant).  One finds that there is a tachyon in the
spectrum of string excitations between the two branes, and this
tachyon causes the system to decay into a stable configuration in
which supersymmetry is restored.

In our system, the two A-type D-branes we start with preserve a
different phase in the A-type linear combination of supercharges and
hence break supersymmetry.  Since they are \slag\ branes, they are
minimal volume elements in their homology class with respect to the
\Kah\ structure of the LSM target space, but their sum is not minimal
volume in its homology class because of the homologically trivial
piece.  There will be a maximally tachyonic mode in the spectrum of
strings stretched between the anti-aligned segments of the D-brane
which will drive a process of tachyon condensation causing these
segments to annihilate to the vacuum.  After annihilation of the
common anti-aligned line segment(s) we are again left with a D-brane
that is minimal volume in its homology class (i.e.~another \slag\ 
brane associated to a rational boundary state).  One can presumably
make this tachyon condensation argument more rigorous in these simple
examples.

In terms of the tachyon condensation picture, when we add together two
D-branes that have {\it both} rays anti-aligned in a coordinate plane
(with the rays in the other coordinates all equal), they will
annihilate completely to give a D-brane that is trivial in this
coordinate.  Topologically, two such D-branes have opposite
orientation, and they are indeed found to have opposite D-brane
charges which therefore cancel to give the vacuum.

\section{New boundary states from geometry}
\label{newbranes}

In section \ref{lg} I constructed the set of \slag\ submanifolds of
$\C^5/\Z_5$ that span the toric base of the orbifold; these are the
``base-filling D-branes'' or ``spokes'' discussed in that section.  As
I have shown, this class of D-brane reproduces the properties of the
rational boundary states of the Gepner model, to which the linear
sigma model flows in the infrared at the point $r \rightarrow -\infty$
in \Kah\ moduli space.

Recall that these submanifolds are characterized by a constant value
of the angular coordinates $\theta_i$ along the submanifold.
Therefore, the \Kah\ form (\ref{kahlercn}) will vanish term by term
upon restriction to the submanifold; i.e.~these submanifolds are
separately Lagrangian with respect to each coordinate $z_i$ of the
target space.

This is directly analogous to the Recknagel-Schomerus construction of
boundary states in Gepner models and its further analysis by Fuchs
et.~al.: recall that this construction preserves an \ntwo\ \SUSY\ 
algebra in each minimal model factor of the Gepner model.  Since
Lagrangian submanifolds correspond to boundary states of CFT that
preserve \ntwo\ {\it worldsheet} \SUSY, the preserved \SUSY\ algebras
of the CFT and geometrical constructions are in direct agreement.

It is then clear how one might proceed to study a more general class
of ``non-rational'' boundary states of the Gepner model, by relaxing
this symmetry condition in the linear sigma model and studying \slag\ 
submanifolds that do not preserve the Lagrangian condition in each
coordinate separately, but only for the expression (\ref{kahlercn}) as
a whole (and which intersect the vacuum submanifold of the LG phase,
i.e.~the origin in $\C^5/\Z_5$).  This symmetry relaxation seems
difficult to study directly in conformal field theory since in general
it renders the theory non-rational, but the mapping of this problem
into geometry shows how it may be approached geometrically.

One large class of more general \slag\ submanifolds are those that
respect the toric structure of the linear sigma model target space,
i.e.~the \slag\ submanifolds with $p \neq 0$ in the notation of
section \ref{lg}.  Recall that they are defined by taking additional
hyperplane constraints in the toric base, and corresponding orthogonal
subspaces of the torus fibre.  Since the \slag s with $p \neq 0$ do
not have fixed values of the angular coordinates $\theta_i$ they meet
the requirement discussed in the previous paragraph.  Note that there
are a countably infinite number of these \slag\ submanifolds since
they are defined by vectors of integer charges (\ref{slaghyper}) that
specify the normal vectors to the hyperplanes.  In general the
translation moduli of these hyperplanes are fixed in the linear sigma
model by the requirement that they intersect the orbifold fixed point
(the vacuum submanifold of the LG phase, where the CFT lives in the
infrared).  At large volume these \slag s possess moduli since they
have $b_1 \neq 0$; these moduli may survive at the Gepner point if
they are not lifted by an instanton-generated superpotential.

Taking hyperplanes defined by normal vectors with {\it non-integral}
entries induce subtori of the $T^n$ fibres that are non-rational,
i.e.~they foliate the $T^n$.  This is a familiar example in
non-commutative geometry and it may be possible to understand these
``non-commutative'' \slag s (and corresponding ``non-commutative
boundary states'') in that context.

There may be other more general classes of \slag\ submanifolds that
can be studied in the linear sigma model framework and which could be
used to define boundary states in CFT: for example \slag\ submanifolds
that do not respect the toric description of the target space, or that
are constructed using more general involutions of the target space
(one such possibility currently under study in CFT is
\cite{permbranes}).

There are no obviously defined notions of geometry within the Gepner
model itself.  However, the structure of the bulk Gepner model as a
conformal field theory (on a worldsheet without boundary) can be
thought of as a remnant of the geometry of the bulk linear sigma model
(i.e.~without D-branes) in this limit.  For example, the
superpotential of the Landau Ginzburg orbifold theory (which generates
the chiral ring) survives as the spectrum of chiral primary operators
of the Gepner model \cite{chiralrings, lgminimal}, and the tensor
product structure of the Gepner model descends from the $\C^5/\Z_5$
geometry of the linear sigma model target space as explained above.

In the same way that the bulk Gepner model retains a ``memory'' of the
bulk linear sigma model, I propose that the boundary states of the
Gepner model should be thought of as remnants of the \slag\ 
submanifolds of the linear sigma model.  The construction of Recknagel
and Schomerus and Fuchs et.~al.~amounts to reconstructing this
``latent geometry'' for the class of $p=0$ \slag s, which are simple
enough to construct from first principles in CFT because of their high
degree of symmetry.

For more general \slag\ submanifolds of the linear sigma model target
space, a corresponding boundary state of the CFT would also be defined
by RG flow, but certain properties could be studied within the linear
sigma model directly, as was done in previous sections for the class
of \slag\ that descend to the rational boundary states in the
infrared.

\section{Conclusions and future directions}
\label{conc}

I have studied the simplest class of toric \slag\ submanifolds of the
target space of a linear sigma model, which descend to A-type D-branes
of string theory on the compact \CY\ in the infrared and which
reproduce the topological properties of the rational boundary states
of the Gepner model.  Some of these submanifolds are only piecewise
\slag, but since they are in the same homology class as a true \slag\ 
we can still use them for topological purposes.  Furthermore, for the
quintic a true \slag\ in this homology class can always be found
explicitly.

For the LG model associated to an \ntwo\ minimal model, the behaviour
of Lagrangian D-branes under variation of complex structure was
considered in \cite{dmirror}.  The decay of \slag\ submanifolds is a
classical geometrical problem which has been studied in \cite{joyce2,
  joyce}, and it is not corrected by instantons in string theory.  It
should be possible to analyze this problem in the linear sigma model,
and it may be easier to study than the equivalent B-type problem,
where destabilization may be caused by instanton effects and the
mathematical description of the decay process is more complicated
\cite{stability, dcat, decat0, dstabmon}.

The discussion in section \ref{chargelatt} about relationships within
the lattice of rational D-brane states has much of the flavour of a
tachyon condensation argument, in which two rational D-brane states
condense to another state (possibly another rational boundary state)
that is of minimal volume in the same total homology class (hence an
A-type D-brane).  It would be interesting to analyze the dynamics of
this process concretely from the point of view of boundary RG flows in
the boundary linear sigma model \cite{horilsm, openlsm, bfermions},
and to use it to study the fate of unstable intersecting rational
boundary states that do not share a common boundary (i.e.~for which
the tachyon is non-maximal), which are expected to decay to a
(presumably non-rational) bound state.

Perhaps the most interesting possibility to emerge from the linear
sigma model construction of A-type D-branes is the construction of new
boundary states of the Gepner model.  It would be interesting to
investigate these possibilities in more detail.
$\\$

{\bf Acknowledgements:} I would like to thank P.~Berglund, I.~Brunner,
N.~Halmagyi, W.~Lerche, A.~Recknagel, C.~R\"{o}melsberger,
C.~Schweigert, J.~Walcher and V.~Yasnov for helpful discussions, and
especially N.~Warner for his advice and guidance during this work.

\appendix
\section{Existence of special Lagrangians in the set of spokes}
\label{slagproof}

{\bf Theorem:} For any pair of half-branes described by $(k+2)^{\rm th}$
roots of unity $\{n_i\}$, $\{\overline{n}_i\}$ in $\C^{k+2}/\Z_{k+2}$,
where $(k+2)$ is a prime, there is a \slag\ submanifold in the same
homology class that is also described by pairs of $(k+2)$ roots of
unity.
$\\$
{\bf Proof:} Performing a field identification (\ref{fieldid}):
\begin{equation}
(L_i, M_i, S_i) \sim (k_i - L_i, M_i + k_i + 2, S_i + 2)
\end{equation}
on some subset of the $i$ labels does not change the homology class of
a D-brane (up to an overall minus sign for an odd number of field
identifications, corresponding to an anti-brane), although it does
change the value of $L_i$ and therefore the grade of the brane (one
can verify this by computing the charges of the D-branes according to
\cite{quintic}).

Geometrically, the field redefinition corresponds to changing the
orientation of the D-brane in one of the coordinates $z_i$, which
interchanges $n_i$ from the set of angles labeling the ``incoming''
ray with $\overline{n}_i$ labeling the ``outgoing'' ray; the new pair
of half-branes is a different choice of cycle in the same homology
class, which has a different grade on each of the half-branes since we
have reassigned the $n_i$.

Suppose the initial grades $G = \sum_{i=1}^{k+2} \theta_i$,
$\overline{G} = \sum_{i=1}^{k+2} \overline{\theta}_i$.  The $\theta_i$
are valued in $U(1)$, but if we are looking for a solution amongst the
$(k+2)^{\rm th}$ roots of unity then we restrict to the $\Z_{k+2}$
subgroup of $U(1)$ generated by the roots of unity, labeled by the
$n_i$.  If $G = \overline{G}$ then the submanifold is already \slag\ 
and we are done.  For $G \neq \overline{G}$ we want to find a subset
$I \subset \{1, \ldots, k+2\}$ such that interchanging $n_i
\leftrightarrow \overline{n}_i$ for each $i \in I$ gives
\begin{equation}
G' = \sum_{i=1}^{k+2} n'_i = \overline{G}' = \sum_{i=1}^{k+2} \overline{n'}_i 
\end{equation}

The interchange operation on $I$ shifts $G$ and $\overline{G}$ by
\begin{eqnarray}
G' &=& G - S \nonumber \\
\overline{G}' &=& \overline{G} + S \nonumber \\
S &=& \sum_{i \in I} \left( n_i - \overline{n}_i \right) \equiv \sum_{i \in I} \Delta n_i
\end{eqnarray}
Therefore
\begin{equation}
G - \overline{G} = 2 S
\label{shiftedg}
\end{equation}
The LHS is equal to $\sum_{i=1}^{k+2} L_i$ and is given.  If ${k+2}$
is even, then there exist elements $G-\overline{G}$ of $\Z_{k+2}$ for
which (\ref{shiftedg}) has no solution for $S$ (namely the odd
elements of $\Z_{k+2}$), because the operation ``division by two'' is
not well-defined on $\Z_{2m}$.  Therefore for $k$ even there is a
subset of the spoke pairs that cannot be brought into \slag\ form by
interchange operations.

Restricting to $k$ odd, the element $S$ exists for all values of
$\sum_i L_i = G - \overline{G}$ and the problem may be reformulated as
follows: find a subset $I \in \{1, \ldots, k+2\}$ such that
\begin{eqnarray}
\sum_{i \in I} \Delta n_i &=& S \nonumber\\
\sum_{i=1}^{k+2} \Delta n_i &=& 2 S = \sum_{i=1}^{k+2} L_i
\end{eqnarray}

If $(k+2)$ is prime, $\Z_{k+2}$ has no proper subgroup, and any $r
\geq k+2$ elements of $\Z_{k+2}$ will generate the entire group by
taking all possible partial sums.  Therefore, for $(k+2)$ prime and $r
\geq k+2$ every choice of $\{L_i\}$ has a set of interchange
operations to bring it into \slag\ form and the theorem is proven.  If
$r < k+2$ then there may exist elements $2S \in \Z_{k+2}$ for which
the corresponding $S$ cannot be obtained by partial sum, because the
$\{\Delta n_i\}$ do not span the entire group under partial summation.

If $\Z_{k+2}$ has a proper subgroup, i.e.~$k+2$ is composite, then the
elements $\{\Delta n_i\}$ may again fail to span the entire group
under partial summation, because they can span the subgroup $H$ and
possibly some of its cosets $g+H$ without spanning the entire group.

Therefore, for cases other than $Z_{k+2}$ prime, $r \geq k+2$ there
exist spoke pairs in the model that are in the same homology class as
a \slag\ submanifold (the existence of the A-type rational boundary
states ensures the existence of such a \slag), but for which that
\slag\ does not exist in the set of spoke pairs.  It is generally only
a subset of spoke pairs that fail to be globally \slag, and there are
also true \slag\ submanifolds in the set of spoke pairs.  $\\\square$

The conditions of this theorem are true for the $(k=1)^3$ and
$(k=3)^5$ Gepner models, which correspond respectively to a torus
$T^2$ with periodicity given by the $su(3)$ root lattice, and the
quintic hypersurface in $\P^4$.


\newpage
\begin{thebibliography}{10}
\bibitem{gepner}D.~Gepner, {\it Exactly Solvable String
    Compactifications On Manifolds Of SU(N) Holonomy},
  Phys.~Lett.~{\bf B199} (1987) 380.
  
\bibitem{gepner2}D.~Gepner, {\it Space-Time Supersymmetry In
    Compactified String Theory And Superconformal Models},
  Nucl.~Phys.~{\bf B296} (1988) 757.

\bibitem{chernsimons}E.~Witten, {\it Chern-Simons Gauge Theory as a
    String Theory}, hep-th/9207094.
  
\bibitem{bbs}K.~Becker, M.~Becker, A.~Strominger, {\it Fivebranes,
    Membranes and Non-perturbative String Theory}, Nucl.\ Phys.\ {\bf
    B456} (1995) 130, hep-th/9507158.
  
\bibitem{ooy}H.~Ooguri, Y.~Oz, Z.~Yin, {\it D-Branes on Calabi-Yau
    Spaces and Their Mirrors}, Nucl.\ Phys.\ {\bf B477} (1996) 407,
  hep-th/9606112.

\bibitem{rs}A.~Recknagel, V.~Schomerus, {\it D-branes in Gepner
    Models}, Nucl.~Phys.~{\bf B531} (1998) 185, hep-th/9712186.
  
\bibitem{bstates}J.~Fuchs, C.~Schweigert, J.~Walcher, {\it Projections
    in String Theory and Boundary States for Gepner Models}, Nucl.\ 
  Phys.\ B {\bf 588} (2000) 110, hep-th/0003298.

\bibitem{dmirror}K.~Hori, A.~Iqbal, C.~Vafa, {\it D-branes and Mirror
    Symmetry}, hep-th/0005247.
  
\bibitem{lgcft}S.~Govindarajan, T.~Jayaraman, {\it On the
    Landau-Ginzburg Description of Boundary CFTs and Special
    Lagrangian Submanifolds}, JHEP {\bf 0007} (2000) 016, hep-th/0003242.
  
\bibitem{dglsm}S.~Govindarajan, T.~Jayaraman, T.~Sarkar, {\it On
    D-branes from Gauged Linear Sigma Models}, Nucl.\ Phys.\ {\bf
    B593} (2001) 155, hep-th/0007075.

\bibitem{quintic}I.~Brunner, M.~Douglas, A.~Lawrence,
  C.~R\"{o}melsberger, {\it D-branes on the Quintic}, JHEP {\bf 0008}
  (2000) 015, hep-th/9906200.
  
\bibitem{delliptic}D.-E.~Diaconescu, C.~R\"{o}melsberger, {\it
    D-branes and Bundles on Elliptic Fibrations}, Nucl.\ Phys.\ {\bf
    B574} (2000) 245, hep-th/9910172.
  
\bibitem{k3fib}P.~Kaste, W.~Lerche, C.~A.~L\"{u}tken, J.~Walcher, {\it
    D-branes on K3-Fibrations}, Nucl.\ Phys. {\bf B582} (2000) 203, hep-th/9912147.

\bibitem{bpsnoncompact}M.~R.~Douglas, B.~Fiol, C.~R\"{o}melsberger,
  {\it The Spectrum of BPS Branes on a Noncompact Calabi-Yau},
  hep-th/0003263.
  
\bibitem{stability}M.~R.~Douglas, B.~Fiol, C.~R\"{o}melsberger, {\it
    Stability and BPS branes}, 
  hep-th/0002037.

\bibitem{dcat}M.~R.~Douglas, {\it D-branes, Categories and \nOne\ 
    Supersymmetry}, hep-th/0011017.

\bibitem{decat0}P.~S.~Aspinwall, A.~Lawrence, {\it Derived Categories
    and Zero-Brane Stability}, JHEP {\bf 0108} (2001) 004, hep-th/0104147.
  
\bibitem{dstabmon}P~S.~Aspinwall, M.~R.~Douglas, {\it D-brane
    Stability and Monodromy}, hep-th/0110071.

\bibitem{joyce2}D.~Joyce, {\it Lectures on Calabi-Yau and special
    Lagrangian Geometry}, math.DG/0108088.

\bibitem{joyce}D.~Joyce, {\it On Counting Special Lagrangian Homology
    3-spheres}, hep-th/9907013.

\bibitem{stringycy}D.-E.~Diaconescu, M.~R.~Douglas, {\it D-branes on
    Stringy Calabi-Yau Manifolds}, hep-th/0006224.

\bibitem{candelas}P.~Candelas, X.~De La Ossa, P.~Green, L.~Parkes,
  {\it A Pair of Calabi-Yau Manifolds as an Exactly Soluble
    Superconformal Theory}, Nucl.\ Phys.\ {\bf B359} (1991) 21.
  
\bibitem{periods}P.~Berglund, E.~Derrick, T.~H\"{u}bsch,
  D.~Jan\v{c}i\'{c}, {\it On Periods for String Compactifications},
  Nucl.\ Phys.\ {\bf B420} (1994) 268, hep-th/9311143.
  
\bibitem{gepnersuper}I.~Brunner, V.~Schomerus, {\it On Superpotentials
    for D-Branes in Gepner Models}, JHEP {\bf 0010} (2000) 016,
  hep-th/0008194.
  
\bibitem{quinticW} M.~R.~Douglas, S.~Govindarajan, T.~Jayaraman,
  A.~Tomasiello, {\it D-Branes on Calabi-Yau Manifolds and
    Superpotentials}, hep-th/0203173.

\bibitem{branestoric}N.~C.~Leung, C.~Vafa, {\it Branes and Toric
    Geometry}, Adv.\ Theor.\ Math.\ Phys.\ {\bf 2} (1998) 91,
  hep-th/9711013.
  
\bibitem{holdiscs}M.~Aganagic, C.~Vafa, {\it Mirror Symmetry, D-Branes
    and Counting Holomorphic Discs}, hep-th/0012041.
  
\bibitem{wittenn2}E.~Witten, {\it Phases of \ntwo\ Theories in Two
    Dimensions}, Nucl.\ Phys.\ {\bf B403} (1993) 159, hep-th/9301042.
  
\bibitem{cardy1}J.~Cardy, {\it Conformal Invariance and Surface
    Critical Behavior}, Nucl.~Phys.~{\bf B240} (1984) 514.
  
\bibitem{cardy2}J.~Cardy, {\it Effect of Boundary Conditions on the
    Operator Content of Two-Dimensional Conformally Invariant
    Theories}, Nucl.~Phys. {\bf B275} (1986) 200.
  
\bibitem{ishibashi1}N.~Ishibashi, {\it The Boundary and Crosscap
    States in Conformal Field Theories}, Mod.~Phys.~Lett. {\bf A4}
  (1989) 251.
  
\bibitem{ishibashi2}N.~Ishibashi, T.~Onogi, {\it Conformal Field
    Theories on Surfaces with Boundaries and Crosscaps},
  Mod.~Phys.~Lett. {\bf A4} (1989) 161.
  
\bibitem{singcurv}I.~Brunner, V.~Schomerus, {\it D-branes at singular
    curves of Calabi-Yau compactifications}, JHEP {\bf 0004} (2000)
  020, hep-th/0001132.

\bibitem{complexman}P.~Candelas, {\it Lectures on Complex Manifolds},
  in {\it Trieste 1987, Proceedings, Superstrings '87}, 1-88.
  
\bibitem{newmanifolds}A.~Strominger, E.~Witten, {\it New Manifolds for
    Superstring Compactification}, Commun.\ Math.\ Phys.\ {\bf 101}
  (1985) 341.
  
\bibitem{mirrorlsm}K.~Hori, C.~Vafa, {\it Mirror Symmetry},
  hep-th/0002222.

\bibitem{greeneplesser}B.~R.~Greene, M.~R.~Plesser, {\it Duality In
  Calabi-Yau Moduli Space}, Nucl.~Phys.~{\bf B338} (1990) 15.

\bibitem{openclosed}P.~Mayr, {\it \nOne\ Mirror Symmetry and
    Open/Closed String Duality}, hep-th/0108229.

\bibitem{openmirror}W.~Lerche, P.~Mayr, {\it On \nOne\ Mirror Symmetry
    for Open Type II Strings}, hep-th/0111113.
  
\bibitem{openinst}S.~Kachru, S.~Katz, A.~E.~Lawrence, J.~McGreevy,
  {\it Open String Instantons and Superpotentials}, Phys.\ Rev.\ D
  {\bf 62} (2000) 026001, hep-th/9912151.
  
\bibitem{mirroropen}S.~Kachru, S.~Katz, A.~E.~Lawrence, J.~McGreevy,
  {\it Mirror Symmetry for Open Strings}, Phys.\ Rev.\ D {\bf 62}
  (2000) 126005, hep-th/0006047.

\bibitem{dualityweb}M.~Aganagic, A.~Klemm, C.~Vafa, {\it Disk
    Instantons, Mirror Symmetry and the Duality Web}, Z.\ 
  Naturforsch.\ {\bf A57} (2002) 1, hep-th/0105045.
  
\bibitem{semiperiods}A.~C.~Avram, E.~Derrick, D.~Jan\v{c}i\'{c}, {\it
    On Semi-Periods}, Nucl.\ Phys.\ {\bf B471} (1996) 293, hep-th/9511152.
  
\bibitem{mclean}R.~McLean, {\it Deformations of Calibrated
    Submanifolds}, Comm.\ Anal.\ Geom.\ {\bf 6} (1998), 705.
  
\bibitem{syz}A.~Strominger, S.~T.~Yau, E.~Zaslow, {\it Mirror symmetry
    is T-duality}, Nucl.\ Phys.\ {\bf B479} (1996) 243,
  hep-th/9606040.

\bibitem{dlectures}C.~Bachas, {\it Lectures on D-branes}, hep-th/9806199.

\bibitem{wound}S.~Mizoguchi, T.~Tani, {\it Wound D-branes in Gepner
    Models}, Nucl.\ Phys.\ {\bf B611} (2001) 253, hep-th/0105174.

\bibitem{sen1}A.~Sen, {\it Tachyon Condensation on the Brane Antibrane
    System}, JHEP {\bf 9808} (1998) 012, hep-th/9805170.
  
\bibitem{sen2}A.~Sen, {\it Descent Relations Among Bosonic D-branes},
  Int.\ J.\ Mod.\ Phys.\ {\bf A14} (1999) 4061, hep-th/9902105.
  
\bibitem{relevance}J.~A.~Harvey, D.~Kutasov, E.~J.~Martinec, {\it On
    the Relevance of Tachyons}, hep-th/0003101.
  
\bibitem{braneantibrane}Y.~Oz, T.~Pantev, D.~Waldram, {\it
    Brane-Antibrane Systems on Calabi-Yau Spaces}, JHEP {\bf 0102}
  (2001) 045, hep-th/0009112.
  
\bibitem{permbranes}A.~Recknagel, {\it Permutation Branes}, work in
  progress.

\bibitem{chiralrings}W.~Lerche, C.~Vafa, N.~P.~Warner, {\it Chiral Rings
    in \ntwo\ Superconformal Theories}, Nucl.~Phys.~{\bf B324} (1989) 427.
  
\bibitem{lgminimal}E.~Witten, {\it On the Landau-Ginzburg Description
    of \ntwo\ Minimal Models}, Int.\ J.\ Mod.\ Phys.\ {\bf A9} (1994)
  4783, hep-th/9304026.

\bibitem{horilsm}K.~Hori, {\it Linear Models of Supersymmetric
    D-branes}, hep-th/0012179.
  
\bibitem{openlsm}S.~Hellerman, S.~Kachru, A.~Lawrence, J.~McGreevy,
  {\it Linear Sigma Models for Open Strings}, hep-th/0109069.
  
\bibitem{bfermions}S.~Govindarajan, T.~Jayaraman, {\it Boundary
    Fermions, Coherent Sheaves and D-branes on Calabi-Yau Manifolds},
  Nucl.\ Phys.\ {\bf B618} (2001) 50, hep-th/0104126.
  
\bibitem{disclsm}S.~Govindarajan, T.~Jayaraman, T.~Sarkar, {\it Disc
    Instantons in Linear Sigma Models}, hep-th/0108234.

\end{thebibliography}
\end{document}